\documentclass[reprint,superscriptaddress,nofootinbib,amsmath,amssymb,aps,prl,onecolumn,notitlepage]{revtex4-1}

\usepackage[a4paper, margin=2.50cm]{geometry}

\usepackage{bbm}
\usepackage{graphicx}
\usepackage{fancyhdr, color}
\usepackage{hyperref} 
\usepackage{tikzsymbols}
 
\hypersetup{
	colorlinks=true,
	linkcolor=blue,
	citecolor=blue,
	urlcolor=blue
}

\definecolor{martin}{rgb}{0,.4,1}

\definecolor{henrik}{rgb}{1,.4,0}

\newcommand{\mc}[1]{\mathcal{#1}}

\newcommand{\mb}[1]{\mathbb{#1}}

\renewcommand{\Re}{\mathrm{Re}}

\newcommand{\tr}{\mathrm{Tr}} 
\newcommand{\Tr}{\mathrm{Tr}} 

\newcommand{\id}{\mathbb{I}}

\newcommand{\mcq}{\mc{Q}^{(\psi)}}

\newcommand{\R}{\mb{R}}

\newcommand{\ket}[1]{|#1\rangle}
\newcommand{\bra}[1]{\langle #1|}

\newcommand{\ketbra}[2]{|#1\rangle\langle #2|}

\newcommand{\dd}{\textrm{d}}

\newcommand{\upsi}{^{(\psi)}}

\newtheorem{thm}{Theorem}

\newtheorem{corol}[thm]{Corollary}
\newtheorem{defn}{Definition}

\begin{document}

\fancyhead[C]{\sc \color[rgb]{0.4,0.2,0.9}{Quantum Thermodynamics book}}
\fancyhead[R]{}

\title{Fluctuating work in coherent quantum systems: proposals and limitations}

\author{Elisa B\"aumer}
\email{ebaeumer@itp.phys.ethz.ch} 
\affiliation{Institute for Theoretical Physics, ETH Zurich, Wolfgang-Pauli-Str. 27, 8093 Z\"urich, Switzerland}

\author{Matteo Lostaglio}
\email{lostaglio@gmail.com} 
\affiliation{ICFO-Institut de Ciencies Fotoniques, The Barcelona Institute of Science and Technology, Castelldefels (Barcelona), 08860, Spain}

\author{Mart\'i Perarnau-Llobet}
\email{marti.perarnau@mpq.mpg.de}
\affiliation{Max-Planck-Institut f\"ur Quantenoptik, Hans-Kopfermann-Str. 1, D-85748 Garching, Germany}

\author{Rui Sampaio}
\email{rui.ferreirasampaio@aalto.fi}
\affiliation{QTF Center of Excellence, Department of Applied Physics, Aalto University, P.O. Box 11000, FI-00076 Aalto, Finland.}

\date{\today}

\begin{abstract}
One of the most important goals in quantum thermodynamics is to demonstrate advantages of thermodynamic protocols over their classical counterparts. For that, it is necessary to (i) develop theoretical tools and experimental set-ups to deal with quantum coherence in thermodynamic contexts, and to (ii)  elucidate which properties are genuinely quantum in a thermodynamic process. In this short review, we discuss proposals to define and measure work fluctuations that allow to capture quantum interference phenomena. We also discuss fundamental limitations arising due to measurement back-action, as well as connections between work distributions and quantum contextuality. We hope the different results summarised here motivate further research on the role of quantum phenomena in thermodynamics.
\end{abstract}

\maketitle

\thispagestyle{fancy}


The aim of this short review is to discuss different proposals and definitions of fluctuating work in quantum systems, particularly in the presence of a superposition of energies in the initial state. This is an exciting topic that has recently received a lot of attention in the community of quantum thermodynamics~\cite{Goold2016,Vinjanampathy2016}. One can identify different motivations behind this research line:
\begin{enumerate}
\item Fluctuations are of the same order of magnitude as average quantities in microscopic systems~\cite{aberg2013truly}. 
As a consequence, the standard laws of thermodynamics, which are formulated for average quantities, are not enough to characterise the thermodynamics of nano-scale systems~\cite{TPM1RMP,Esposito2009nonequilibrium,erratumm,jarzynski2011equalities}.

\item Fluctuations can be of an intrinsic nature in quantum systems, since due to quantum coherence there are states of maximal knowledge (pure states) for which our ability to predict the outcome of an energy measurement is very limited.
As we are mostly concerned with energy considerations, we use the term ``quantum coherence" to express that a quantum state is in a superposition of different energy eigenstates  (a ``coherent state", not to be confused with the standard notion from quantum optics). 
 
\item Recent results suggest that quantum coherence and entanglement play a role in several thermodynamic tasks, such as heat engines \cite{Rossnagel2014nanoscale,Correa2014quantum,Alicki2015Noneq,
Brask2015Small,uzdin2015quantum,
Mitchison2015Coherence,Hofer2016autonomous,Nimmrichter2017andclassical,
Brandner2017universal,Klatzow2017Experimental} and work extraction \cite{Allahverdyan2004maximal,Funo2013work,mpl2015extractable,korzekwa2016extraction,
EnergycostPRA2016,Patrick2018}, in some cases leading to potential enhancements  \cite{Karen2013,Brunner2014entanglement,Raam2015equivalence,Campaioli2017,
Ferraro2017,watanabe2017quantum,Patrick2018,levy2018quantum}. 
\end{enumerate} 
It turns out, however, that even the definition of  fluctuating work (and similarly heat) in the presence of quantum coherence is a challenging task. Roughly speaking, this is because thermodynamic quantities, such as heat and work, are associated to a process, rather than to a single time event.  
Hence, in order to deal with fluctuations, one wishes to build a stochastic theory for processes. 
In classical physics, this can be naturally done by identifying trajectories in phase space. However, given a quantum evolution in time, there is no  unique notion of a quantum trajectory. 
 A quantum trajectory can be built by measuring the system at different times. Nevertheless, the measurement of an observable affects the statistics of non-commuting observables one may need to access later \cite{busch2014quantum}. 

The goal of this short review is to elaborate more on these subtle points, and to discuss some proposals that give different insights into the notion of quantum work. We are concerned with questions like: What is work in the quantum regime? Can we define and measure fluctuations in processes with quantum interference? Can the quantum measurement back-action be reduced? Let us start by giving a bit of context to these questions.

In classical systems, fluctuations of thermodynamic quantities have been studied extensively, leading to seminal results in the form of {\it Fluctuation Theorems} (FTs) (see e.g. \cite{Jarzynskia2008} for a review). 
 First attempts to extend FTs to the quantum regime were based on the identification of work as an observable (i.e. a Hermitian operator) \cite{Bochkov1977,BOCHKOV1981443,Yukawa2000,Monnai2003,Chernyak2004,
Allahverdyan2005,Engel2007,Gelin2008}, leading to corrections to the standard FTs. It was later recognized that the standard  FTs could be obtained by defining fluctuating work as the difference between the outcomes of two projective energy measurements (TPM) \cite{TPM3arxiv,TPM4arxiv,TPM2PRE,TPM1RMP,Esposito2009nonequilibrium,
Hanggi2015theother,Campisi2011Colloquim,Campisi2011ColloquimII}. The TPM scheme  gives  a clear operational and physical meaning to work, is applicable to both closed and open systems \cite{Campisi2009}, has been implemented in different experimental set-ups \cite{Huber2008employing,Tiago2014experimental,Shuoming2014experimental,
Cerisola2017quantum}, and one can define a natural correspondence with the classical definition of work  \cite{Jarzynski2015quantum,Zhu2016quantum}. This makes the TPM scheme  a standard  definition of fluctuating work in quantum systems nowadays.  

Despite its great success, recently several authors have pointed out limitations of the TPM scheme, especially when applied to initial coherent states \cite{NQFoWPRE2014,Solinas15,Kammerlander2016,Deffner2016Quantum,NoGoTheorem}. Some of the most relevant ones are:
\begin{enumerate}
\item Projective energy measurements are difficult to implement in certain experimental set-ups. This has motivated alternative measurements of work based on noisy energy measurements \cite{GEMPhysRevE,QFTaPMNJP2015}. 
\item The quantum non-equilibrium free energy \cite{EspositoSecondLaw} can be additively decomposed into an incoherent and a coherent contribution \cite{janzing2006quantum, lostaglio2015description}. An energy measurement sets the coherent contribution to zero, hence changing the non-equilibrium properties, \emph{e.g.} decreasing the maximal average amount of extractable work  (see Appendix). 
\item  More generally, the TPM scheme is invasive and can perturb the state it acts upon.   In particular, the first energy measurement 
projects the initial state onto the energy eigenbasis, thus destroying possible off-diagonal terms. This prevents the possibility of capturing purely coherent evolutions, i.e., interference effects due to the initial coherence in the state. 
\end{enumerate}
It is important to note that these limitations affect only states with coherence, and that similar limitations appear in other measurement schemes; after all, all quantum measurements are invasive in some way, have a thermodynamic cost, and quantum systems are challenging to prepare, control and measure. At the same time, one hopes that complementary proposals for measuring work can capture new aspects and provide new insights in out-of-equilibrium situations. For example, recent proposals study the possibility of using weak measurements to define quantum trajectories \cite{QFTaPMNJP2015,Alonso2016thermodynamics,Elouard2017role}. 

In this review we focus on point 3., and discuss alternative proposals for defining and measuring work associated with purely coherent evolutions. These include Gaussian measurements \cite{roncaglia2014work, dechiara2015measuring, Talkner2016aspects}, weak measurements and quasiprobabilities \cite{NQFoWPRE2014,Solinas15,Solinas2017measurement,Hofer2017quasiprobability},  collective measurements \cite{NoGoTheorem} and Bohmian work \cite{sampaio17}.  
We discuss specific proposals in the light of a recent no-go theorem that sets fundamental limitations on the possibility to design (non-invasive) work measurements \cite{NoGoTheorem}, as well as connections between work distributions and quantum contextuality \cite{Lostaglio2018}. We use these no-go results as classification tools.
Finally, we briefly discuss the possibility of defining (fluctuating) work by the state of an external system that acts as a work repository, which has been put forward  in resource-theoretic approaches to thermodynamics \cite{Alhambra2016,richens2016work,Aberg2018,nelly2018quantum} (see also  Ref.~\cite{watanabe2017quantum} for an implementation of a heat engine with a work repository).

\section{Classical work}

We start by briefly discussing the notion of fluctuating work in classical systems.
Given the microscopic state of the system $z = (q,p)$, i.e., its position and momentum, the energy is defined to be the value that the Hamiltonian $ \mathcal{H}(z,t) $ takes for this state at time $t$. The work done on a system evolving under Hamilton's equations can then be defined \footnote{
	Depending on the system and quantity being measured one may use other work definitions. For a concise review see, {\it e.g.}, Ref. \cite{Campisi2011Colloquim}. For our considerations, however, this choice is inconsequential.} as the power integrated along the trajectory $z_t$ followed by the system from time $t=0$ to $t=\tau$, {\it i.e.}, 
\begin{equation}
w = \int_{0}^{\tau} dt \hspace{1mm} \partial_t \mathcal{H} (z_t) = \mathcal{H}(z_\tau,\tau) - \mathcal{H}(z_0,0),
\label{classical1}
\end{equation} 
where $ z_\tau $ and $ z_0 $ are the system's final and initial state, respectively. Hence, under closed ( Hamiltonian) dynamics, the work is just the total change in the energy. The work distribution is then defined to be 
\begin{align}
p_{\rm clas.}(w) := \int\dd z_0 \hspace{1mm} p(z_0) \delta(w - (\mathcal{H}(z_\tau,\tau) - \mathcal{H}(z_0,0))),
\label{classical2}
\end{align}  
where $ p(z_0) $ is the initial distribution of position and momentum. This distribution is well-defined at all intermediate times of the evolution.

\section{Quantum work: Setting the stage}

When dealing with quantum systems, we extend the classical picture by considering a thermally isolated system, initially prepared in  a quantum state $\rho$ and with an internal Hamiltonian $H(0)$. The Hamiltonian can be externally driven leading to a time-dependent Hamiltonian $H(t)$ between time $0$ and $\tau$. We set
\begin{equation}
H(0)=:H = \sum_i E_i \ketbra{E_i }{E_i }, \quad H(\tau)=:H' = \sum_j E'_j \ketbra{E'_j }{E'_j }.
\end{equation}
 This leads to a unitary evolution of the system $\rho(t)=U(t)\rho U^{\dagger}(t)$ with $U(t)= \mathcal{T} \exp \left(-i \int_{0}^t ds H(s) \right)$ and $\mathcal{T}$ the time-ordering operator. We are interested in the energy fluctuations between $0$ and $\tau$, induced by the unitary evolution $U := U(\tau)$. Given a protocol $\mathcal{P}$ (including the specification of $\rho$ and $U$) we will define by $p(w|\mathcal{P})$ the corresponding work distribution. The choice $\mathcal{P}=\mathcal{P}_{\rm TPM}$ of the TPM scheme corresponds to the following steps:
\begin{enumerate}
	\item Projective measurement of $H$ on $\rho$, returning outcome $E_i$ and post-measurement state $\ketbra{E_i}{E_i}$.
	\item Unitary evolution $\ketbra{E_i}{E_i} \mapsto U \ketbra{E_i}{E_i} U^\dag $.
	\item Projective measurement of $H'$ on $U \ketbra{E_i}{E_i} U^\dag$, returning outcome $E'_j$.
\end{enumerate}
The values of work $w^{(ij)}$ and the associated probabilities $p^{(ij)}$ are then given by
\begin{align}
w^{(ij)}=E_j' - E_i, \quad \quad p^{(ij)}_{\rm TPM}=\bra{E_i}\rho\ket{E_i} \left| \bra{E'_j}U  \ket{E_i} \right|^2.
\label{valuesW}
\end{align}
Finally, in order to account for possible degeneracies in the work values $w^{(ij)}$, the work probability distribution is constructed as
\begin{align}
p(w|\mathcal{P}_{\rm TPM})= \sum_{ij} p^{(ij)} \delta\left(w- w^{(ij)} \right)
\end{align}
Note the close analogy with the classical definitions \eqref{classical1} and \eqref{classical2}.  Note also that this scheme becomes invasive when $[\rho,H] \neq 0$ and $[H,U^\dag H' U]\neq 0$. In this case, the statistics collected from the measurement of $H'$ at time $t=\tau$ does not return the same statistics one would have obtained if $H$ was not measured at time $t=0$. In particular, $\sum_w w \, p(w|\mathcal{P}_{\rm TPM})$ in general is not equal to the average energy change induced by $U$ on $\rho$.

\section{Limitations, no-go theorems, and contextuality}

In this section, we describe a no-go result from Ref.~\cite{NoGoTheorem} which sets fundamental limitations on the possibility to design (non-invasive) work measurements. Rather than focusing on a particular approach, the idea is to consider an arbitrary measurement scheme $\mathcal{P}$ to measure the amount of work necessary to perform $U$, and then show that three general requirements cannot be simultaneously satisfied. Later, this  result will serve as a basis to classify different approaches to measure quantum work.   

The three  requirements put forward in Ref.~\cite{NoGoTheorem} read:

\begin{enumerate}
\item The first requirement can be imposed in two equivalent ways (see Appendix~A of Ref.~\cite{NoGoTheorem}): 
 \label{convexitypositivity}
\begin{enumerate}
 \item \emph{$p(w|\mathcal{P})$ is a linear  probability distribution.}
 This corresponds to the assumption that $p(w|\mathcal{P})$ is a probability distribution $p(w|\mathcal{P})>0$,  $ \sum p(w|\mathcal{P})=1$, linear under mixtures of $\rho$. Linearity is defined as: Let $\mathcal{P}^{i}$, with $i=0,1,2$, be protocols differing only by the initial state $\rho_i$. Then, if $\rho_0= \lambda \rho_1 + (1-\lambda) \rho_2$ with $\lambda \in [0,1]$, we demand that $p(w|\mathcal{P}^0)=\lambda  p(w|\mathcal{P}^1) + (1-\lambda) p(w|\mathcal{P}^2)$.  This corresponds to the natural requirement that, if we condition 
the choice of the protocol on a coin toss, the measured
fluctuations are simply the convex combination of
those observed in the individual protocols.

\item There exists a Positive-Operator-Valued Measure (POVM), i.e., a set of positive operators $\{\Pi_w\}$, dependent on  $H$, $H'$ and $U$ but \emph{not} $\rho$, that satisfy $p(w|\mathcal{P})= \tr(\rho \Pi_w)$ and $\sum_w \Pi_w = \id$. 
\end{enumerate}
 To understand the relation between (a) and (b), note that if (b) is not satisfied, then one can apply different measurements to $\rho_1$, $\rho_2$ and $\lambda \rho_1+(1-\lambda) \rho_2$; so that the corresponding work distribution will not in general be convex. For a formal proof of their equivalence, we refer the reader to the first section of the Supplemental Material of Ref.~\cite{NoGoTheorem}. 

 \item \emph{Agreement with the TPM scheme for non-coherent states.} \label{agreement} The second requirement is based on the assumption that the TPM scheme yields the correct statistics for diagonal states, i.e.
 \begin{align}
p(w|\mathcal{P}) = p(w|\mathcal{P}_{\rm TPM}) \hspace{15mm} \forall \rho^{\rm} \;\; \textrm{    such that   } \; \; [\rho^{\rm}, H] = 0.
 \end{align}
  This condition is motivated by the success of the TPM scheme in describing work fluctuations for diagonal states, in particular in fluctuation theorems, and by results concerning the recovery of the classical limit \cite{Jarzynski2015quantum}.
 
 \item \label{average} \emph{Average energy changes are respected by the measurement process.} Finally, we demand that the average energy change predicted by $\mathcal{P}$ equals the average energy change induced by $U$ on $\rho$:
\begin{align}
\label{eq:unmeasuredwork}
\sum_w w \hspace{1mm} p(w|\mathcal{P}) = \tr(U\rho U^{\dagger} H') - \tr (\rho H) \hspace*{15mm} \forall \rho.
\end{align}
Note that this is imposed for all states, including coherent states. With this requirement, we are essentially demanding that the measurement back action is small, so  that average work is not modified by the measurement scheme used to build the work probability distribution. Eq.~\eqref{eq:unmeasuredwork} can be understood as the first law of thermodynamics, when applied to an isolated system undergoing the evolution $U$.
\end{enumerate}
Note that in a general protocol $\mathcal{P}$ the values of work $w$ in $p(w|\mathcal{P})$ do not need to correspond to $w^{(ij)}$ of Eq.~\eqref{valuesW} and 
can be possibly continuous. We have
\begin{thm}[\cite{NoGoTheorem} No-go for $\mathcal{P}$]
	\label{thm:nogo1}
	There exists no protocol $\mathcal{P}$ satisfying \ref{convexitypositivity}, \ref{agreement} and \ref{average} for all $\rho$,  $U$, $H$, $H'$.
\end{thm}
 This result shows the inherent difficulty to construct a universal scheme to measure work applicable to all processes and  states, including coherent ones. The TPM scheme, for example, satisfies assumptions~\ref{convexitypositivity} and \ref{agreement}, but \ref{average} is satisfied only if $[\rho,H]=0$ and/or $[H,UH'U^\dag] = 0$.
  It also suggests that the subtleties of quantum work fluctuations cannot be captured by a single measurement scheme (or definition), but rather  different measurement schemes reveal different aspects of quantum work. In the next section, 
 we discuss different approaches to the measurement of work fluctuations from the perspective of this no-go result, but before we  discuss its relation with quantum contextuality.

\subsection{ Relation to contextuality }

A different angle from which we can investigate general protocols $\mathcal{P}$ is in terms of the measurement statistics they can generate. The core question is the following: can the statistics collected by the thermodynamic experiment at hand be reproduced through a classical mechanism?

There are different ways of defining what we mean by \emph{classical mechanism}. Clearly, the more liberal this notion, the higher the bar that we are setting to call a phenomenon \emph{non-classical}.
For example, one may call non-classical any system whose state is in a superposition of some preferred eigenstates (\emph{e.g.}, position or energy) or any entangled state -- notions, however, not independent of quantum theory. Another choice would be to call non-classical any phenomenon that is not explained by classical mechanics and the classical theory of radiation. Here, we take a much more liberal approach and define non-classical any phenomenology that cannot be explained within a \emph{non-contextual} model.

Non-contextuality is a notion of classically dating back to the famous no-go theorem of Kochen and Specker \cite{bell1966problem, kochen1975problem}, and used in recent years to identify the origin of quantum advantages in certain models of computation \cite{ howard2014contextuality, delfosse2015wigner, bermejo2016contextuality}. Here we use a rather broad notion of contextuality due to Spekkens \cite{spekkens2005contextuality}.
Before we give a precise claim, this is in summary the relation between contextuality and Theorem~\ref{thm:nogo1}~\cite{Lostaglio2018}:
\begin{itemize}
	\item A protocol $\mathcal{P}$ that reproduces the TPM scheme for diagonal states (assumption \ref{agreement} holds) can probe non-contextuality \emph{only if} $p(w|\mathcal{P})$ is a quasi-probability or lacks convexity (assumption \ref{convexitypositivity} is violated).
	\item There exists a weak measurement protocol satisfying assumptions \ref{agreement} and \ref{average}, with $p(w|\mathcal{P})$ linear in $\rho$, such that \mbox{$p(w|\mathcal{P})<0$} for some $w$ is \emph{sufficient} to prove contextuality.
\end{itemize}
The first point is a rephrasing of Theorem~\ref{thm:nogo1}, and clarifies the role quasi-probabilities play within fluctuation theorems. The second point will be discussed together with alternative proposals later in the text.

\subsubsection{An alternative formulation of the no-go theorem}

Schematically, an experiment boils down to a set of instructions telling us how to set up the system, i.e. a \emph{preparation procedure} $P$, and a set of instructions telling us how to implement a measurement on the system, i.e. a \emph{measurement procedure} $M$. The measurement procedure $M$ returns outcomes $k$ when performed after the preparation procedure $P$, so that the experiment collects the statistics $p(k|P,M)$. 

Two preparation procedures $P$ and $P'$ may be indistinguishable in terms of the statistics they produce, in the sense that $p(k|P,M) = p(k|P',M)$ for every $M$ and $k$. We denote this fact by $P \sim P'$. Similarly, two measurement procedures $M$ and $M'$ are indistinguishable when $p(k|P,M) = p(k|P,M')$ for every $k$ and every preparation procedure $P$, a fact that we denote by $M \sim M'$.

The observed statistics $p(k|P,M)$ is explained by positing the existence of an underlying set of physical states $\lambda$, drawn from a set $\Lambda$ and such that
\begin{itemize}
	\item Every time we follow the instructions $P$, we sample $\Lambda$ according to a distribution $p(\lambda|P)$;
	\item Every time we measure according to $M$, the device returns outcome $k$ with probability $p(k|M,\lambda)$ if the physical state is $\lambda$.
\end{itemize}
The observed statistics is explained as $p(k|P,M)= \sum_{\lambda \in \Lambda} p(\lambda|P) p(k|M,\lambda)$.
Such a model is called an \emph{ontological model} or a \emph{hidden variable model}.
Most famously, certain statistics are incompatible with the assumption that $\lambda$ behaves locally \cite{brunner2014bell}. Non-contextuality, instead, is defined as follows:
\begin{defn}\cite{spekkens2005contextuality}
	An ontological model is called \emph{non-contextual} if for any $P,P'$ with $P \sim P'$ one has $p(\lambda|P) \equiv p(\lambda|P')$ and for any $M,M'$ with $M \sim M'$ one has $p(k|M,\lambda)\equiv p(k|M',\lambda)$.
\end{defn}
In other words, non-contextuality posits that the reason why certain preparations and measurements cannot be operationally distinguished is that they are the same at the ontological level.
While both locality and non-contextuality hold in any classical theory, it is known that they are violated by quantum experiments. Given these definitions, Theorem~\ref{thm:nogo1} gives:
\begin{corol}\cite{Lostaglio2018}
	\label{corol:no-go}
 Assume no degeneracies. Any protocol $\mathcal{P}$ satisfying \ref{convexitypositivity} and \ref{agreement} admits a non-contextual ontological model for the preparation of $\rho$ and the measurement of $p(w|\mathcal{P})$.
\end{corol}
This can be seen as a corollary of Theorem~\ref{thm:nogo1}; in fact, Theorem~\ref{thm:nogo1} is proven by showing that assumptions \ref{convexitypositivity} and \ref{agreement} imply that the POVM $\{\Pi_w\}$ must be the TPM POVM, and the latter is easily seen to admit a non-contextual model. 
This result leads us naturally to review alternative proposals in which $p(w|\mathcal{P})$ is a work quasi-probability distribution or lacks linearity.

\section{ Definitions and measurements of quantum work fluctuations}
\label{Sec:Def}

In this section, we discuss several proposals for definitions of fluctuating work (some of them based on explicit measurement schemes) beyond the standard TPM scheme.  There are protocols that define work probabilities (condition~\ref{convexitypositivity} is satisfied). As imposed by Theorem~\ref{thm:nogo1}, either the average work does not match the average energy change of the unperturbed process, or the distribution disagrees with the TPM one for diagonal states. Some protocols provide nice tradeoffs between these two extremes. Another class of protocols does not satisfy assumption~\ref{convexitypositivity}, either because they lead to a probability distribution that is non-linear or because they yield quasi-probabilities. These can satisfy both conditions~\ref{agreement} and \ref{average}, and Corollary~\ref{corol:no-go} suggests some of them may be linked to contextuality. To give an overview, the different approaches described below are summarized in Table \ref{table1}, where one can see which of the conditions are satisfied in each case.
\begin{table}[h]
\centering
\begin{tabular}{| l | c | c | c |}
\hline
\ Approach & \ Condition 1 \ & \ Condition 2 \ & \ Condition 3 \ \\
\hline
\ TPM \cite{TPM3arxiv,TPM4arxiv,TPM2PRE,TPM1RMP,Esposito2009nonequilibrium,
Hanggi2015theother,Campisi2011Colloquim,Campisi2011ColloquimII} & \Smiley[][green!30!yellow] & \Smiley[][green!30!yellow] & \Sadey[][orange!70!yellow] \\
\ Operator of work \cite{Bochkov1977,BOCHKOV1981443,Yukawa2000,Monnai2003,Chernyak2004,
Allahverdyan2005,Engel2007,Gelin2008} & \Smiley[][green!30!yellow] & \Sadey[][orange!70!yellow] & \Smiley[][green!30!yellow] \\
\ Gaussian measurements \cite{roncaglia2014work, dechiara2015measuring, Talkner2016aspects} & \Smiley[][green!30!yellow] & \Sey[][yellow] & \Sey[][yellow]\\
\ Full-Counting statistics \cite{Solinas15, Solinas2016probing,xu2017effects} & \Sadey[][orange!70!yellow] & \Smiley[][green!30!yellow] & \Smiley[][green!30!yellow]\\
\ Post-selection schemes \cite{aharonov1988result,Lostaglio2018} & \Sey[][yellow] & \Smiley[][green!30!yellow] & \Sey[][yellow]\\
\ Weak values quasi-probability \cite{NQFoWPRE2014,aharonov1988result, wiseman2002weak,Miller2017time,hall2004prior} & \Sadey[][orange!70!yellow] & \Smiley[][green!30!yellow] & \Smiley[][green!30!yellow]\\
\ Consistent Histories \cite{Miller2017time,griffiths1984consistent,goldstein1995linearly} &  \Sadey[][orange!70!yellow] & \Sadey[][orange!70!yellow] & \Smiley[][green!30!yellow]\\
\ Quantum Hamilton-Jacobi \cite{sampaio17}  & \Sadey[][orange!70!yellow] & \Sadey[][orange!70!yellow] & \Smiley[][green!30!yellow] \\
\ POVM depending on the initial state \cite{SAGAWA2012} & \Sadey[][orange!70!yellow] & \Smiley[][green!30!yellow] & \Smiley[][green!30!yellow]\\

\hline
\end{tabular}
\caption{This table summarizes the different approaches described in the following section except for the collective measurement, where the no-go result needs to be adapted. \ \Smiley[][green!30!yellow] \ corresponds to ``satisfied" and \ \Sadey[][orange!70!yellow] \ to ``not satisfied". The list also includes families of protocols in which either one of two conditions is satisfied in a specific limit, with the family interpolating between the two (\ \Smiley[][green!30!yellow] \Sadey[][orange!70!yellow] \ $\rightarrow$ \ \Sadey[][orange!70!yellow] \Smiley[][green!30!yellow] \ ). We denote this situation simply with two \ \Sey[][yellow] \ . We can see that there is no approach satisfying all three conditions, as anticipated by Theorem \ref{thm:nogo1}.}
\label{table1}
\end{table}

\subsection{Operator of work}

The operator of work was one of the first ways proposed to define quantum work fluctuations \cite{Bochkov1977,BOCHKOV1981443,Yukawa2000,Monnai2003,Chernyak2004,
Allahverdyan2005,Engel2007,Gelin2008}. 
The basic idea is to take as a starting point Eq.~\eqref{eq:unmeasuredwork}, write the right hand side as $\tr (\rho (U^{\dagger}H' U-H))$, and define the work operator as
\begin{equation}
\hat{W}=U^{\dagger}H' U-H.
\end{equation} 
This operator is Hermitian, and hence can be diagonalised as $\hat{W}=\sum_i \tilde{w}_i \Pi_i$.
The eigenvalues $\tilde{w}_i$ are then identified with work values, and the corresponding probability reads $\tr(\rho \Pi_i)$. One then defines $p_{OW}(w) = \sum_i \tr{}{[\rho \Pi_i]} \delta(w- \tilde{w}_i)$.

By definition, the operator of work satisfies both conditions \ref{convexitypositivity} and \ref{average}. However, it fails to satisfy condition~\ref{agreement}. This is easy to see, e.g., just note that the work values $\tilde{w}_i$ do not correspond to energy differences of the form $E'_j-E_i$, where $E_i$, $E'_j$ are eigenvalues of $H$ and $H'$, respectively. In fact, while the operator of work is formally and operationally well defined, the interpretation of its fluctuations as work fluctuations is problematic. To illustrate this, suppose that  $\hat W$  has a zero eigenvalue and  consider the corresponding eigenstate $ \ket{\psi}$ with $\hat W \ket{\psi}=0$ as an  initial state $\rho = \ketbra{\psi}{\psi}$. One can still have $\tr{}{[U \rho U^\dag H'^{m}]} \neq \tr{}{[\rho H^{m}]}$ for some $m>2$ and $U$. In other words, the energy distribution of the state can vary (implying some energy exchange) even if $p_{OW}(w) =0$ for $w \neq 0$, i.e. even if there are no fluctuations according to the operator of work \cite{Allahverdyan2014nonequilibrium}.

\subsection{Gaussian measurement}

To explore other schemes, it is important to investigate more carefully the role of the measurement apparatus. Many proposals interpolate between a regime in which the system interacts strongly with the measurement device, recovering the TPM distribution, and one in which the interaction is minimally invasive \cite{roncaglia2014work, dechiara2015measuring, Talkner2016aspects}.

In Refs.~\cite{roncaglia2014work, dechiara2015measuring, Talkner2016aspects}, the projective measurements of the TPM scheme are replaced by a von Neumann measurement scheme in which the Hamiltonian is coupled to the position of a pointer state, also called a ``work meter'' (see Ref. \cite{de2018ancilla} for more details). 
The pointer is taken to be a one-dimensional continuous system with conjugate variables $X$ and $P$, in the simplest case a pure Gaussian state initially prepared with $\langle X \rangle = \langle P \rangle = 0$ and trivial covariance between $X$ and $P$. A measurement is described by the interaction $V = e^{i g H \otimes P}$ (with $g$ the effective interaction strength), 
followed by a projective measurement of the position $X$ of the pointer. Each outcome $x$ defines a corresponding POVM element $M_x$ on the system, corresponding to a more or less invasive measurement depending on $g$ and the initial spread of the pointer. Since $V$ shifts the pointer position by $g E_i$, upon observing outcome $M_x$ one can make the unbiased estimate of the system energy $E=x/g$.
The scheme then consists of: von Neumann interaction $V=e^{i g H \otimes P}$; unitary evolution $U$ on the system; von Neumann interaction $V'=e^{-i g H' \otimes P}$ with the same pointer; measurement of $X$ on the pointer, which defines a work probability distribution $p(w) = p(g x)$ (an alternative scheme adds a position measurement after the first system-pointer interaction \cite{Talkner2016aspects}). 

This protocol satisfies condition~\ref{agreement} only in the strong-measurement limit, when $g$ is large/the pointer distribution is sharp; in this limit, the final measurement destroys all information about the coherences in the initial state (which can be seen by direct calculation of the distribution in Ref.~\cite{roncaglia2014work}) and, in fact, $p(w)$ coincides with the TPM distribution (see Ref. \cite{de2018ancilla} for a similar discussion). For general coupling strength and a diagonal input, the work distribution $p(w)$ is a convolution of a Gaussian (whose width depends on the width of the pointer) and the TPM distribution (Eq. 37 of Ref.~\cite{Talkner2016aspects}), i.e., essentially a ``smeared'' version of the TPM distribution. Conversely, in the weak-measurement limit (where $g$ is small/the pointer distribution is broad), the back action is suppressed and condition~\ref{average} is satisfied (eq. 47 of Ref.~\cite{Talkner2016aspects}); however, condition \ref{agreement} is not.  Furthermore, in this regime work fluctuations diverge as the measurement is completely imprecise \cite{Talkner2016aspects}. We see Theorem~\ref{thm:nogo1} in action in the mutual incompatibility between \ref{average} and \ref{agreement}, given \ref{convexitypositivity}.

\subsection{Full Counting Statistics}

The above considerations lead to a different class of proposals. For general states and in the weak measurement limit, 
$p(w)$ is a convolution of a Gaussian and a function $q(w)$, i.e. a ``smeared'' version of $q(w)$ (eq. 55 of Ref.~\cite{Talkner2016aspects}):
\begin{equation}
\small
q(w) = \sum_{n,n',m} \bra{E_n}\rho \ket{E_{n'}} \tr{}{[\ketbra{E_n}{E_{n'}}  U^\dag \ketbra{E'_m}{E'_m}U]}\hspace{1mm} \delta\hspace{-0.5mm}\left(w- \left(E'_m- \frac{E_n + E_{n'}}{2}\right) \right),
\end{equation}
where $\delta(x-y) =0$ unless $x=y$, in which case it equals $1$. The characteristic function of $q(w)$ can be accessed by the same scheme described for $p(w)$, with the difference one measures the relative phase accumulated by momentum eigenstates of the pointer rather than its position (an approach known as Full Counting Statistics, see Ref.~\cite{Solinas15, Solinas2016probing,xu2017effects}). This approach is rooted in the Keldysh formalism and, ultimately, in the idea of assigning a joint distribution to non-commuting observables \cite{ Hofer2017quasiprobability}. 

$q(w)$ has several interesting properties, and in Ref.~\cite{Solinas2016probing} it was proposed as a work quasi-probability. In general, $q(w)$ equals the TPM distribution plus corrections that depend on the coherence of the initial state and are zero otherwise (Eq. 57 of Ref.~\cite{Talkner2016aspects}).
 Hence, for diagonal states $q(w)$ equals the TPM distribution and condition \ref{agreement} is satisfied. Furthermore, $q(w)$ satisfies condition \ref{average}.  As we know from Theorem~\ref{thm:nogo1}, condition~\ref{convexitypositivity} cannot hold. Hence, since $q(w)$ is linear, it must be a work quasi-probability. Further considerations on this scheme can be found in Ref.~\cite{Solinas2017measurement}, and see also Ref.~\cite{xu2017effects} for discussions on the role of coherence.

\subsection{ Post-selection and Weak values }

Another proposal to assign a joint distribution to $H$ and $U^\dag H' U$ can be obtained from a modification of the scheme described for $p(w)$. Specifically, one can define a work quasi-probability as the weak value of $\ketbra{E_k}{E_k}$ with pre-selection $\rho$ and post-selection $\ket{E'_m}$, where $w= E'_m - E_k$ \cite{aharonov1988result}. Given a pure Gaussian pointer state with spread $s$, the scheme is as follows: 1) the first pointer interaction is $V_k = e^{i g \ketbra{E_k}{E_k} \otimes P}$; 2) Measurement of $X$ on the pointer, which defines a POVM $M^{(k)}_x$ on the system; 3) unitary evolution $U$ on the system; 4) Projective measurement of $H'$ on the system. For more details see Ref.~\cite{Lostaglio2018}, Supplemental Material Sec.~B. Consider now the probability $p(E'_m)$ of observing outcome $E'_m$ in the final measurement,  independently of the observed pointer position. Multiply this by the average position of the pointer given that $m$ is observed, denoted by $\langle X \rangle_{|E'_{m}}$. On the one hand, if step (1) is a strong measurement, $p(E'_m) \langle X \rangle_{|E'_{m}}$ gives the TPM probability {$p^{(km)}_{\rm TPM}$ of Eq.~\eqref{valuesW} (note in this limit the intuition of the quantity $\langle X \rangle_{|E'_{m}}$ as a conditional probability for the initial energy being $k$ given that the final energy is $m$: $\langle X \rangle_{|E'_{m}} \rightarrow |\bra{E'_m}U \ket{E_k}|^2 \bra{E_k}\rho \ket{E_k} / p(E'_m) $ \cite{Lostaglio2018})}. On the other hand, if step (1) is a weak measurement, $p(E'_m) \langle X \rangle_{|E'_{m}}$ returns the Margenau-Hill work quasi-probability $p_{MH}(k,m)$:
\begin{equation}
 p_{MH}(k,m) :=  \Re \tr{}{[ \rho \ketbra{E_k}{E_k} U^\dag \ketbra{E'_m}{E'_m} U]}.
\end{equation}
where  $ \Re (x)$ takes the real part of $x$. One then defines $p_{MH}(w) := \sum_{k,m} p_{MH}(k,m) \hspace{1mm} \delta \hspace{-0.1mm} (w - (E'_{m} - E_k))$. This was first put forward as a work quasi-probability by Allahverdyan \cite{NQFoWPRE2014}, and from the above scheme one can recognize it as the (generalized) weak value \cite{aharonov1988result, wiseman2002weak} of $\ketbra{E_k}{E_k}$ with post-selection $U^\dag \ketbra{E'_m}{E'_m} U$. $p_{MH}(w)$ can also be derived from a consistent histories approach \cite{Miller2017time} and can be understood as an ``optimal'' estimate of the initial energy given the observation of the final energy and the knowledge of the initial state \cite{hall2004prior}. 
The marginals of $p_{MH}$ correspond to the energy distributions of $H$ and $U^\dag H' U$ on $\rho$, so condition~\ref{average} is satisfied. $p_{MH}$ also coincides with the TPM distribution for diagonal states, hence it satisfies condition~\ref{agreement}. By Theorem~\ref{thm:nogo1}, condition~\ref{convexitypositivity} cannot hold, and indeed $p_{MH}$ can become negative \cite{NQFoWPRE2014}. Finally, note that since the variance of $X$ diverges in the weak measurement limit, this scheme requires repeated measurements on a large number of copies of $\rho$. Furthermore, the interaction unitary $V_k$ has to be changed to access $p_{MH}(w)$ for various values of $w$.

\subsubsection{Contextualy proofs from weak values work quasi-probability}

Corollary~\ref{corol:no-go} tells us that among the family of distributions satisfying condition~\ref{agreement}, negativity or more precisely the failure of condition~\ref{convexitypositivity} is a prerequisite to witness contextuality. Is it also sufficient? For the work quasi-probability $p_{MH}(w)$, the following theorem gives a positive answer to this question:  

\begin{thm}\cite{Lostaglio2018}
	 If $p_{MH}(k,m)<0$ for some $E'_{m}$, $ E_k$, if the pointer has spread $s$ large enough there exists no non-contextual ontological model for preparation of $\rho$, measurement of $M^{(k)}_x$, unitary evolution $U$ and post-selection of $\ketbra{E'_m}{E'_m}$. 
\end{thm}
The theorem clarifies that negative values of the work quasi-probability $p_{MH}$ are proofs of contextuality and hence witness non-classicality. This is due to the fact that these negative values are a generalisation of the so-called anomalous weak values, and the latter are known to be proofs of contextuality \cite{pusey2014anomalous}. It is an open question if other work quasi-probabilities are associated to proofs of contextuality.
Finally, we note that anomalous weak values have also been connected to violations of Leggett-Garg inequalities \cite{Williams2008,Solinas15,blattmann2017macroscopic,Miller2017Legget}.

\subsection{Consistent Histories }

Recently, Miller and Anders \cite{Miller2017time} proposed a time-reversal symmetric work distribution based on the Consistent Histories (CH) interpretation of quantum mechanics \cite{griffiths1984consistent,goldstein1995linearly}. In this approach, work is associated with the histories of the power operator in Heisenberg picture $ X_\text{H}(t) \equiv U^\dagger(t)\partial_t H(t) U(t)$, where recall that $ H(t) $ is the Hamiltonian of the system at time $t$ and $ U(t) $ is the evolution operator from time $0$ to $ t $. The Hilbert space is taken to be finite of dimension $ d $. Time is discretized into a grid of K+1 equally spaced points $ \{0,t_1,\dots,t_j,\dots,t_K\} $ and at each time point the spectral decomposition of $ X_\text{H}(t_j) = \sum_{n=1}^d x_n^{(j)} P_n^{(j)} $ is used to build a trajectory $ \vec{n} = (x_{n_0},\dots,x_{n_j},\dots,x_{n_K}) $. 
Here, $ \{P_n^{(j)}\}_{n=1}^d $ is a complete set of orthonormal projectors decomposing $ X_\text{H}(t_j) $, with the corresponding eigenvalues $ \{x_{n}^{(j)}\}_{n=1}^d $. 
Each trajectory $ \vec{n} $ is associated with a class operator $ C_{\vec{n}} = \overset{\leftarrow}{\mathcal{T}} \prod_{j=0}^{K} P_{n_j}^j $ which defines the trajectory quasi-probability $ p_{\vec{n}} \equiv \tr\{(C_{\vec{n}}^\dagger + C_{\vec{n}})\rho\}/2 $, where $ \rho $ is the initial state and $ \overset{\leftarrow}{\mathcal{T}} $ is the time-ordering operator, arranging the terms from right to left with increasing time. The work for a given trajectory $ \vec{n} $ is then defined to be $w_{\vec{n}} = \sum_{j=1}^{K-1} x_{n_j}^{(j)} \Delta t  $, where $ \Delta t $ is the time step of the grid. 
Finally, the work quasi-probability $ p_{CH}(w) $ is given by grouping the trajectories into {\it histories}, such that, $ p_{CH}(w) = \tr\{(C_{w}^\dagger + C_{w})\rho\}/2 $, where $ C_w = \sum_{\vec{n}} C_{\vec{n} }\delta(w - w_{\vec{n}})  $ and $ \delta(x-y) $ is equal to one if $x=y$ and zero otherwise.

The distribution respects the average energy change of the unperturbed process (requirement 3) but, in general, does not agree with the TPM scheme for initial diagonal states. 
In addition, the work distribution admits a notion of time-reversal symmetry, namely it respects $ \tilde{p}_\mathrm{CH}(w)  = p_\mathrm{CH}(-w)$, where $\tilde{p}_\mathrm{CH}(w)$ is the ``time-reversed work distribution'', constructed from the time-reversed process. Unitary processes are time-symmetric and it may be argued that the work distribution should reflect this symmetry \cite{Miller2017time}.
In the present case, $\tilde{p}_\mathrm{CH}(w)$ is described by reversing the trajectory $ \vec{n} $, which leads to the time-reversed class operator $ \tilde{C}_{w} = C^\dagger_{-w} $. It then follows that $ \tilde{p}_\mathrm{CH}(-w) = p_\mathrm{CH}(w)$.
 As a final remark, in the continuous limit $K \rightarrow +\infty$, we have that $\sum_w w^k p_\mathrm{CH}(w)={\rm Tr}(\rho \hat{W}^k)$ where $\hat{W}$ is the work operator. However,  note that $p_\mathrm{CH}(w)$ and $p_\mathrm{OW}(w)$ can be very different, as $p_\mathrm{CH}(w)$ is in general a quasi-probability distribution and it is non-zero over a different set of $w$ \cite{Miller2017time}.

\subsection{ Hamilton-Jacobi equation} 

 Consider a pure state $\psi(q,t) = R^{(\psi)}(q,t) e^{i S^{(\psi)}(q,t)/\hbar}$. Another alternative \cite{sampaio17} is based on the Hamilton-Jacobi formulation of the Schr\"{o}dinger equation, namely, 
\begin{align}
-\partial_t S\upsi(q,t) & =
\frac{{p\upsi}(q,t)^2}{2m} + V(q,t) + \mcq(q,t); \label{eq:qhj-hj}\\
\partial_t \rho\upsi(q,t) & = - \nabla j\upsi(q,t) \label{eq:qhj-cont},
\end{align}
where $S\upsi(q,t)/\hbar \in \R$ is the overall phase of the state $ \psi $, $p\upsi(q,t) \equiv \nabla S\upsi(q,t)$, $\rho\upsi(q,t) \equiv R\upsi(q,t)^2$ is the probability density, $j\upsi(q,t) \equiv \rho\upsi \nabla S\upsi/m $ is the probability current density and $\mcq \equiv -\hbar^2\nabla R\upsi/2mR\upsi$ is the so-called quantum potential. Since Eq.~\eqref{eq:qhj-hj} is an Hamilton-Jacobi equation, one identifies the energy as $ E\upsi(q,t) \equiv -\partial_t S\upsi(q,t) $. One can formally construct a Hamiltonian function $ \mathcal{H}\upsi(q,p,t) = p^2/2m + V(q,t) + \mathcal{Q}\upsi(q,t) $ that generates the flow lines in phase space $ \dot{q} = \partial_p \mathcal{H} $ and $ \dot{p} = - \partial_q \mathcal{H} $. Thus, Eqs.~(\ref{eq:qhj-hj}) and (\ref{eq:qhj-cont}) describe the evolution of a statistical ensemble of point particles in phase, which is closely connected to the Bohmian \cite{Bohm:1952aa,Bohm:1952ab} or Pilot-Wave \cite{broglie1956tentative} interpretations of quantum mechanics. A notable difference from the classical case is the single-valuedness of the $ S $ function and that all quantities are fully determined in configuration space, since the momentum of a particle at position $ q $ in a state $ \psi(q,t) $ is fixed to be $ p\upsi(q,t) $. In other words, while in classical mechanics the state is completely determined by specifying position and momentum, in Bohmian mechanics, the state is specified by the position {\it and} wave function. A detailed description can be found in textbooks, {\it e.g.}, Refs.~\cite{holland1995quantum,durr2012quantum}. 

For unitary evolution, work can then be defined in a manner fully analogous to the classical case (see Introduction), namely, $ W\upsi[q_\tau] = \mathcal{H}\upsi(q_\tau,\tau) - \mathcal{H}\upsi(q_0,0) $. If the system is initially prepared in a state $ \psi $, the work distribution is simply $ P(W;\psi) = \int\dd q |\psi(q,0)|^2 \delta(W - W\upsi[q_\tau]) $. If the system is explicitly prepared in a statistical mixture of pure states \footnote{We take the distribution to be discrete for simplicity sake} $ \{ \psi_j \}_{j=1}^N $ with distribution $ \{p_j\}_{j=1}^{N} $, $ p_j > 0 $, $ \sum_{j=1}^{N} p_j = 1 $, the work distribution is just the convex sum of the individual distribution, {\it i.e.}, $ P(W) = \sum_{j=1}^{N} p_j P(W;\psi_j) $. 

The work distribution is positive and respects the average energy change of the unperturbed process (requirement 3).
Agreement with the TPM scheme is verified when the measurement steps do not affect the state significantly, see Ref.~\cite{sampaio17} for details. 
The main conceptual difference from the previous approaches, apart from the phase-space description, is that it is defined for wave functions rather than density operators. As a consequence, the work is explicitly dependent on the way the system is physically prepared or measured, that is, on the experimental details of the implementation of the protocol. In Ref.~\cite{sampaio17} an example is given for which different statistical mixtures of pure states corresponding to the same density operator give rise to different work distributions. These are experimentally accessible via weak measurement schemes~\cite{AAV88,Duck89,Kocsis2011,Mahler2016,Xiao2017} if a decomposition is explicitly selected, e.g. by a corresponding measurement that separates $\rho$ into pure state components, as also discussed in Ref.~\cite{Allahverdyan2005}.
	If the system is initially entangled with other degrees of freedom, then the work distribution can be experimentally accessed only through measurements on the overall pure state, which might not be known if all the empirically accessible information is that of the reduced density operator. Nonetheless, one may still define a coarse-grained work based on the available information, but this falls outside the scope of this review.
		 Similarly to other approaches not satisfying condition~\ref{agreement}, it is an interesting open question how and if one can recover the appropriate classical limit.

Although quantitatively different, this approach shares many qualitative similarities to a ``sub-ensemble'' approach, as proposed in Ref.~\cite{Allahverdyan2005}. There, the initial state described by a density operator $\rho$ is first split into pure states ensembles
$ \{\ketbra{\psi_\alpha}{\psi_\alpha}\}_{\alpha=1}^N $, with probabilities $ \{\lambda_\alpha\}_{\alpha=1}^N $, such that $ \rho = \sum_{\alpha=1}^N \lambda_\alpha \ketbra{\psi_\alpha}{\psi_\alpha} $.  The fluctuating work is taken to be the change in the energy expectation value for each element of the sub-ensemble. Similar properties include, {\it e.g.}, positive work distribution, average work equal to unperturbed average energy change (condition~\ref{average}), ensemble dependence and non-convexity under mixtures of $ \rho $.  The main difference is that from the Hamilton-Jacobi perspective the individual states $ \ketbra{\psi_\alpha}{\psi_\alpha} $ are treated as representing another sub-ensemble of point-particles trajectories. From this perspective, the sub-ensemble approach can be viewed as a coarse-graining of the Hamilton-Jacobi approach.

\subsection{POVM depending on the initial state}
Another logical possibility is to look at measurement schemes that depend on the initial state. We have seen before that the two-projective-measurement scheme (TPM) does not satisfy condition~\ref{average} on non-diagonal states.
This is because the TPM scheme projects onto the basis of the Hamiltonian and thereby destroys all coherence, so that the state to which the unitary is applied is not the initial one.
However, if the initial state is known, one can always determine the basis in which it is diagonal and conduct a projective measurement onto that basis. Then, no coherence is destroyed during the first measurement and a probability distribution that satisfies both conditions \ref{agreement} and \ref{average} can be obtained. However, the dependence on the initial state directly violates condition~\ref{convexitypositivity}, in accordance to Theorem~\ref{thm:nogo1}. In particular, the resulting probability distribution is not convex. We note that this kind of measurement schemes have been formally considered to obtain general fluctuation theorems \cite{SAGAWA2012}; however, when the initial $\rho$ does not commute with the Hamiltonian the physical meaning of the statistics becomes unclear \cite{SAGAWA2012}.  In particular, there is no clear recipe to assign an energy or work cost  to a transition between two pure states that are in coherent superpositions of energies.

\subsection{Collective measurements}
Another approach, aiming at reducing the measurement back-action of the TPM scheme, is to apply a collective measurement on $N$ copies of the state, each of them undergoing the same unitary evolution. This possibility has been considered in Ref.~\cite{NoGoTheorem}.  Since the measurement acts on $N$ copies, it is convex on the whole state $\rho^{\otimes N}$, but it is not anymore convex at the local level of $\rho$. For global measurements, one can naturally adapt the conditions of the no-go theorem as \cite{NoGoTheorem}: 

\begin{enumerate}
	\item \emph{POVM.} There exists a Positive-Operator-Valued Measure (POVM), i.e., a set of positive operators $\Pi_w$, dependent on $H$, $H'$ and $U$ but \emph{not} $\rho$, that satisfies $p(w|\mathcal{P})= \tr(\rho^{\otimes N}\Pi_w)$ and $\sum_w \Pi_w = \id$. Note that the ${\Pi_w}$ in this case are matrices with dimension $d^N \times d^N$, where $d$ is the local dimension of $\rho$. This gives a much bigger freedom to the measurement scheme. 
	\item \emph{Agreement with the TPM scheme for non-coherent states.} \label{agreementglobal} The $N$-copies protocol recovers the single-copy TPM scheme for diagonal states, i.e., $\tr(\rho_{\rm diag}^{\otimes N} \Pi^{(i)})= \tr(\rho_{\rm diag} \Pi_{\rm TPM}^{(i)}) \quad \forall \; \rho_{\rm diag} $
	such that  $ [\rho_{\rm diag}, H] = 0$.
	\item \emph{Average energy changes are respected by the measurement process.} The $N$-copies average energy change equals the single-copy average energy change induced by $U$ on $\rho$:
	\begin{align}
	\label{eq:averageglobal}
	\sum_w w  \tr(\rho^{\otimes N}\Pi_w) = \tr(U\rho U^{\dagger} H) - \tr (\rho H) \hspace*{15mm} \forall \rho.
	\end{align}
\end{enumerate}
It was shown in Ref.~\cite{NoGoTheorem} that it is still impossible to satisfy these weaker requirements simultaneously. 
That is, even when considering global measurements, there is an intrinsic back-action in measurements of quantum work.  Nevertheless, global measurements allow for improvements, in the sense that one can design measurements on $\rho^{\otimes N}$ that satisfy 1. and 2. exactly, and lead to an average energy that is closer to satisfying Eq.~\eqref{eq:averageglobal} as compared to the single-copy case. In particular, the following collective measurement scheme on two copies of the state has been suggested in \cite{NoGoTheorem}:
\begin{align}
M^{(ij)}_{\lambda}= \ketbra{i}{i} \otimes \left(\bra{i} T_j \ket{i}   \id + \lambda T_{j}^{\rm off-diag} \right) ,
\label{M^{(ij)}}
\end{align}
where
$T_{j}=U^{\dagger} \ketbra{j'}{j'} U$, $T_{j}^{\rm off-diag}$ is the off-diagonal part of $T_j$ in the basis of $H$,  and $\lambda \in [0,1]$ is a parameter that can depend on the specific $U$ \cite{NoGoTheorem}.
The $M^{(ij)}_{\lambda}$ are POVM elements acting on two copies of the state, so that $P^{(ij)}_{\lambda}=\tr{(M^{(ij)}_{\lambda}\rho^{\otimes 2})}$ is the probability assigned to the work cost $w^{(ij)}$ as given in \eqref{valuesW}.  This collective measurement scheme has some appealing properties: 
\begin{itemize}
	\item It satisfies conditions 1. and 2. exactly. The first condition is satisfied by noting that \eqref{M^{(ij)}} are POVM elements, and the second one   follows from  $\tr(\rho^{\rm diag} T_{j}^{\rm off-diag})=0$.
	\item The second term of  \eqref{M^{(ij)}} brings information about the purely coherent part of the evolution, hence extending the standard TPM scheme to partly describe coherent evolutions.
	\item The global measurement \eqref{M^{(ij)}} can be expressed as two independent measurements on each copy of $\rho$, specifically a projective energy measurement on the first copy, and a non-projective measurement on the second copy that depends on $U$. By measuring each copy separately, one before and one after the evolution, the measurement back-action can be reduced.
\end{itemize}
These properties make collective measurements an interesting possibility to  extend the TPM scheme to capture some of the subtle effects of coherent evolutions. Notably, the collective measurement \eqref{M^{(ij)}} has been recently  implemented experimentally in a photonic set-up~\cite{kangda2018}. 

In the classical limit $N \rightarrow \infty$ one can get arbitrarily close to satisfying all requirements; in fact, taking $H_N = \sum_{i} H^{(i)}/N$ ($X^{(i)}$ denotes $X$ in the Hilbert space of particle $i$ and identity elsewhere) and $\tilde{H}'_N = \sum_{i} U^{\dag (i)} H^{'(i)}U^{(i)} /N$, one gets $[H_N, \tilde{H}'_N] \propto 1/N$; in other words, since all average observables commute in the asymptotic limit, and since $\tr(U\rho U^{\dagger} H) - \tr (\rho H) = \tr(\rho ^{\otimes N} \tilde{H}'_N) - \tr (\rho^{\otimes N} H_N)$, non-commutativity becomes increasingly irrelevant: the TPM scheme satisfies all three assumptions arbitrarily well in the limit. However, this does not solve the issue for the thermodynamics of a small number of quantum systems.

\subsection{Beyond work distributions}

A new promising approach developed a fully quantised version of the protocols described above \cite{Aberg2018, Alhambra2016}. Specifically, one can treat the work reservoir as a quantum system, rather than assume a classical external system. Furthermore, one can explicitly model a switch responsible for the change in the system Hamiltonian from $H$ to $H'$ (the overall Hamiltonian is time independent, but the switch induces an effective time-dependent Hamiltonian on the system). One can then impose that the overall dynamics is unitary and conserves energy. In this way the work reservoir, as well as providing the energy necessary for the transformation, also behaves as a quantum probe for the system dynamics. 

For states that are initially thermal, fluctuation-like theorems can then be written for the quantum channel induced on the work source, a quantum version of the standard theorems for $p(w|\mathcal{P})$. These can be tested by process tomography. The standard fluctuation theorems are recovered by looking at the evolution of the diagonal part of the work reservoir, under an additional assumption, i.e. energy translation invariance of the dynamics of the work reservoir. This assumption captures the idea that only displacements in energy of the work reservoir should matter for the definition of work. A Jarzynski equality can be verified by a TPM scheme on the work reservoir, thanks to the decoupling of diagonal and off-diagonal dynamics of the work reservoir (which follows from overall energy conservation and the assumption that the initial state of the system is thermal). Further relations hold for the evolution of the off-diagonal elements of the work reservoir. These results, and an extension to arbitrary initial states described through conditional fluctuation theorems \cite{Aberg2018}, may provide a new avenue to the study of work fluctuations for coherent initial states. For further information, see also \cite{nelly2018quantum}.

\section{Outlook}

The standard approach to measure work fluctuations, consisting of two projective energy measurements (TPM) before and after the evolution, becomes invasive when the initial state is coherent. This prevents the possibility of capturing interference effects arising due to the coherence in the initial state.
This observation has motivated several proposals for measurements (or definitions) of work beyond the standard TPM scheme \cite{Solinas15,Talkner2016aspects,NoGoTheorem,Lostaglio2018,sampaio17,Aberg2018,
	Alhambra2016,Miller2017time,NQFoWPRE2014,
	Solinas2017measurement,Hofer2017quasiprobability}.
The goal of this short review has been to put together these works in order to gain a global perspective of the topic.

As discussed in Ref.~\cite{NoGoTheorem}, the three desiderata (1) Convexity and existence of a work probability distribution, (2) Recovery of the TPM work distribution for non coherent input states, (3) Respecting the identification of average work as average energy change in the unitary process, are mutually incompatible. Furthermore, (1) and (2) alone are incompatible with the possibility of witnessing contextuality \cite{Lostaglio2018}, a distinctive property of quantum phenomena that has been connected to quantum advantages in the realm of computation \cite{howard2014contextuality, delfosse2015wigner}. Here, we have used these impossibility results to understand and classify the different definitions of work fluctuations in coherent evolutions, even though we also mentioned some notable proposals that are not captured within this framework \cite{Aberg2018, Alhambra2016}.

The first conclusion one can draw is that while classically there is a single definition of work for closed system evolutions, there are multiple relevant definitions of work for quantum systems (as also discussed in Ref. \cite{de2018ancilla}). Rather than focusing on the question of what is \emph{the} quantum work, our view is that the relevant questions should be more pragmatic. For a given scheme, what information is encoded in the corresponding work (quasi) probability distribution? What  results exist connecting the features of such quantum work distribution to the underlying properties of the quantum process? 
 Can the obtained statistics be simulated in classical systems (see Refs.~\cite{pashayan2015estimating,blattmann2017macroscopic,Lostaglio2018})? If they cannot, can this lead to any genuine quantum advantage in thermodynamic tasks? Finally, can one construct fluctuation theorems for out-of-equilibrium coherent states (see \cite{NQFoWPRE2014,Solinas2017measurement,Aberg2018} for first attempts) and, more importantly, what is their physical meaning? We hope the results discussed here provide a solid starting point  to  address these questions, and motivate further research in this direction.
  
While many partial results appeared in the literature pointing at potential quantum advantages \cite{Karen2013,Brunner2014entanglement,Raam2015equivalence,Campaioli2017,
Ferraro2017,watanabe2017quantum,Patrick2018,Amikam2018}, proving a clear-cut departure from the classical world view is a necessary condition to identify any truly quantum advantage. What do we mean here by ``truly''? We mean an advantage proved under very weak modelling assumptions, such as locality in Bell inequalities \cite{pironio2010random} or certain operational equivalences for contextuality \cite{mazurek2016experimental}. This allows to pinpoint genuine quantum phenomena through a rigorous no-go theorem, as apparently quantum behaviours can often be reproduced by classical models, also in thermodynamic contexts \cite{Nimmrichter2017andclassical}.   

While showing that a phenomenon defies classical explanations in a broad sense is a necessary condition for any truly quantum advantage, it is by no means sufficient. Here, quantum work distributions could provide a useful bridge, since on the one hand they may be related to thermodynamically relevant quantities, while  on the other hand they may be associated to non-classicality proofs \cite{Solinas15,Miller2017Legget,blattmann2017macroscopic,Lostaglio2018}. It is in this role of probes of the behaviour of quantum systems that quantum work distributions may be most useful in helping us to identify a provable advantage of a thermodynamic protocol over its classical counterparts. Ultimately, this remains one of the most important questions posed by quantum thermodynamics.  \\ \\ \\

\appendix

\section{Appendix: Loss in maximal amount of average extractable work due to energy measurement}

\label{appendix}

The maximal average amount of work that can be extracted from a quantum system in state $\rho$ with internal Hamiltonian $H_S$ and unitary operations on systems and a thermal bath at temperature $T$ is
\begin{equation}
\label{eq:maxaverage}
\langle W \rangle_{max}(\rho) = F(\rho,H_S) - F(\tau,H_S) \equiv kT S(\rho||\tau), 
\end{equation}
where $\tau = \frac{e^{-\beta H_S}}{Z}$, $Z = \tr{}{[e^{-\beta H_S}]}$, $\beta = 1/(kT)$ with $k$ Boltzmann constant, $F(X,H) = \tr{}{[H X]} - kT S(X)$ with $S(X):= - \tr{}{[X \log X]}$ the von Neumann entropy, $S(X||Y):= \tr{}{[X (\log X - \log Y)]}$ the quantum relative entropy. 

The result \eqref{eq:maxaverage} can be inferred from, \emph{e.g.}, the derivations in \cite{EspositoSecondLaw,skrzypczyk2014work,aberg2014catalytic,Reeb2014}. However, for clarity of the exposition, we present here a simple proof. Consider a bipartite system $SB$ with an initial non-interacting  Hamiltonian
\begin{equation}
H=H_{\rm S}+H_{\rm B},
\end{equation}
 prepared in a product state
\begin{equation}
\rho_{SB}=\rho_{\rm S}\otimes \rho_{\rm B},
\label{initialState}
\end{equation}
with $B$ a Gibbs state at inverse temperature $\beta = (kT)^{-1}$: 
\begin{equation}
\rho_{\rm B}=\tau_{B}=\frac{e^{-\beta H_{\rm B}}}{\mathcal{Z}}.
\end{equation}
Hence we can think of $B$ as an auxiliary thermal state. Note that $B$ is not necessarily a thermal bath, in the sense that it may not be macroscopic. We now consider a generic closed evolution in $SB$, which can always be described as a unitary operation $U$:
\begin{equation}
\rho'_{SB}=U \rho_S \otimes  \rho_B U^{\dagger},
\end{equation}
The average extracted work is given by the energy change on $SB$,
\begin{equation}
W=\Tr[H \rho_{SB}]-\Tr[H \rho'_{SB}] = F(\rho_{SB},H) - F(\rho'_{SB},H),
\end{equation}
where we assumed that at the end of the interaction the final Hamiltonian is again $H$, and for the second equality we used $S(\rho'_{SB}) = S(\rho_{SB})$. 

For any bipartite system $X_{SB}$ with Hamiltonian $H$, denoting by $X_{S/B} = \Tr_{B/S}X_{SB}$, one has that the non-equilibrium free energy decomposes into local parts plus correlations:
\begin{equation}
F(X_{SB}, H) = F(X_S,H_S) + F(X_B,H_B) + kT I(X_{SB}),
\end{equation}
where $I(X_{SB}) = S(X_S) + S(X_B) - S(X_{SB})$ is the mutual information (this can be verified by summing and subtracting the local entropies to the expression for $F(X_{SB}, H)$). 

Denoting now $\rho'_{S/B}=\Tr_{B/S}[\rho'_{SB}]$ and using the formula above, together with the fact that $I(\rho_{SB}) = 0$, we obtain
\begin{align}
W&=F(\rho_S,H_S)-F(\rho'_S,H_S) +F(\rho_B,H_B)-F(\rho'_B,H_B) -kT I(\rho'_{SB})
\nonumber\\
&=F(\rho_S,H_S)-F(\rho'_S,H_S) -kT (S(\rho'_{B}||\tau_B) +I(\rho'_{SB})),
\label{WorkGeneralII}
\end{align}
where we used $F(\rho'_B,H_B)-F(\tau_B,H_B)=kT S(\gamma_B||\tau_{B})$.
As both $I(\cdot)$ and $S(\cdot||\cdot)$ are non-negative quantities, we immediately obtain 
\begin{equation}
W\leq  F(\rho_S,H_S)-F(\rho'_S,H_S).
\end{equation}  
Note that this expression only depends on the (initial and final) state of $S$ and the temperature of $B$. We can obtain a bound that is independent of the final state by adding and subtracting $F(\tau_S,H_S)$ in Eq.~\eqref{WorkGeneralII} (where $
\tau_{\rm S}=e^{-\beta H_S}/\Tr(e^{-\beta H_S})$), which gives
\begin{equation}
W= F(\rho_{\rm S},H_S)-F(\tau_{\rm S},H_S)-T\left[ S(\rho'_S||\tau_{\rm S})+I(\rho'_{SB})+S(\rho'_B||\tau_{\rm B}) \right]
\label{WorkGeneral},
\end{equation}
Note again that all terms in the square parenthesis are non-negative, and each of them has an intuitive physical meaning: in order of appearance, the athermality of the final state of $S$ (when $\rho'_S\neq \tau_S$), the correlations created between $S$ and $B$, and the athermality of final state of $B$ (when $\rho'_B \neq \tau_B$). With the above equality we finally obtain
\begin{equation}
W\leq F(\rho_{S},H_S)-F(\tau_{S},H_S) = \langle W \rangle_{max}(\rho).
\end{equation}
The remaining interesting question is whether these bounds can be saturated: protocols that achieve \eqref{eq:maxaverage} are constructed in \cite{skrzypczyk2014work,aberg2014catalytic,Reeb2014} which, interestingly, require $B$ to be of macroscopic size.

We now move to the question of how energy measurements change Eq.~\eqref{eq:maxaverage}, by making it unattainable. For simplicity of exposition, we will assume that $H_S$ is not degenerate. If one performs an energy measurement, one obtains a pure energy state $E_i$ with probability $p_i = \bra{E_i} \rho \ket{E_i}$. Using Eq.~\eqref{eq:maxaverage}, one can see that from state $\ket{E_i}$ one can extract a maximum amount of work equal to $ F(\ket{E_i}) - F(\tau) = E_i + kT \log Z$ (this can be understood as the conjunction of the unitary process that maps $\ket{E_i}$ into the ground state, extracting energy $E_i$, followed by a protocol that extracts work $kT \log Z$ from the purity of the ground state). To complete the process one needs to reset the memory, implicitly used in the measurement, to its ``blank state''; in the presence of a bath at temperature $T$, this requires an investment of $kT H(\vec{p})$, where $\vec{p}$ is the distribution $p_i$ and $H(\vec{p}) = - \sum_i p_i \log p_i$ is the Shannon entropy (this is known as Landauer erasure). Overall the protocol that extracts work after an energy measurement optimally achieves the average $\langle W \rangle_{meas}(\rho) = \sum_i p_i E_i  - kT H(\vec{p}) + kT \log Z$. A direct calculation shows $\langle W \rangle_{meas} =  F(\mathcal{D}(\rho)) - F(\tau)$, with $\mathcal{D}$ the dephasing operation $\mathcal{D}(\rho) = \sum_i p_i \ketbra{E_i}{E_i}$. This implies 
\begin{equation}
\langle W \rangle_{meas}(\rho) = \langle W \rangle_{max}(\mathcal{D}(\rho)),
\end{equation}
i.e. the energy measurement protocol optimally extracts an average amount of work equal to the maximum that can be extracted by first dephasing the state and then performing work extraction. This implies a loss 
\begin{equation}
\langle W \rangle_{max}(\rho) - \langle W \rangle_{max}(\mathcal{D}(\rho)) = kT S(\mathcal{D}(\rho)) - kT S(\rho) \equiv kT S(\rho||\mathcal{D}(\rho)):= kT A(\rho),
\end{equation}
proportional to a quantity $A(\rho)$ called \emph{asymmetry} or \emph{relative entropy of coherence}, which is a measure of quantum coherence in the eigenbasis of $H_S$. Note that $A(\rho) > 0$ if and only if $\rho \neq \mathcal{D}(\rho)$.

The above reasoning shows that protocols based on energy measurements cannot reach the optimal average work extraction, since they lose the possibility of extracting work from the coherence of the quantum state. This intuition can also be grounded in the non equilibrium free energy. As we discussed above, $\langle W \rangle_{max}(\rho) = \Delta F (\rho) := F(\rho) - F(\tau)$. Summing and subtracting $\Delta F(\mathcal{D}(\rho)) = F(\mathcal{D}(\rho)) - F(\tau)$ one obtains, using the definition of $A(\rho)$,
\begin{equation}
\Delta F(\rho) = \Delta F(\mathcal{D}(\rho)) + kT A(\rho),
\end{equation}
i.e. the non-equilibrium free energy neatly decomposes into a contribution coming from the diagonal part of the state and a contribution coming from coherence \cite{janzing2006quantum, lostaglio2015description}. As discussed above, $\langle W \rangle_{meas} =  \Delta F(\mathcal{D}(\rho))$, i.e. the diagonal non-equilibrium free energy captures the component that can be converted into work by the energy measurement protocol, whereas the coherent contribution is lost. One has $\Delta F(\rho)>\Delta F(\mathcal{D}(\rho))$ whenever $[\rho,H_S]\neq 0$.

 \bigskip
 
\emph{Acknowledgements}.  We thank  Johan \AA{}berg, Armen Allahverdyan, Janet Anders, Peter H\"anggi, Simone Gasparinetti, Paolo Solinas and Peter Talkner for useful feedback on the manuscript.  E.B. acknowledges contributions from the Swiss National Science Foundation via the NCCR QSIT as well as project No.\ 200020\_165843. M.P.-L. acknowledges support from the Alexander von Humboldt Foundation.  M.L. acknowledges financial support from the the European Union's Marie Sklodowska-Curie individual Fellowships (H2020-MSCA-IF-2017, GA794842), Spanish MINECO (Severo OchoaSEV-2015-0522 and project QIBEQI FIS2016-80773-P), Fundacio Cellex and Generalitat de Catalunya (CERCAProgramme and SGR 875).
R. S. acknowledges  the Magnus Ehrnrooth Foundation and the Academy of Finland through its CoE grants 284621 and 287750. All authors are grateful for support from the EU COST Action MP1209 on Thermodynamics in the Quantum Regime. 

\bibliography{Review}

\begin{thebibliography}{116}%
\makeatletter
\providecommand \@ifxundefined [1]{%
 \@ifx{#1\undefined}
}%
\providecommand \@ifnum [1]{%
 \ifnum #1\expandafter \@firstoftwo
 \else \expandafter \@secondoftwo
 \fi
}%
\providecommand \@ifx [1]{%
 \ifx #1\expandafter \@firstoftwo
 \else \expandafter \@secondoftwo
 \fi
}%
\providecommand \natexlab [1]{#1}%
\providecommand \enquote  [1]{``#1''}%
\providecommand \bibnamefont  [1]{#1}%
\providecommand \bibfnamefont [1]{#1}%
\providecommand \citenamefont [1]{#1}%
\providecommand \href@noop [0]{\@secondoftwo}%
\providecommand \href [0]{\begingroup \@sanitize@url \@href}%
\providecommand \@href[1]{\@@startlink{#1}\@@href}%
\providecommand \@@href[1]{\endgroup#1\@@endlink}%
\providecommand \@sanitize@url [0]{\catcode `\\12\catcode `\$12\catcode
  `\&12\catcode `\#12\catcode `\^12\catcode `\_12\catcode `\%12\relax}%
\providecommand \@@startlink[1]{}%
\providecommand \@@endlink[0]{}%
\providecommand \url  [0]{\begingroup\@sanitize@url \@url }%
\providecommand \@url [1]{\endgroup\@href {#1}{\urlprefix }}%
\providecommand \urlprefix  [0]{URL }%
\providecommand \Eprint [0]{\href }%
\providecommand \doibase [0]{http://dx.doi.org/}%
\providecommand \selectlanguage [0]{\@gobble}%
\providecommand \bibinfo  [0]{\@secondoftwo}%
\providecommand \bibfield  [0]{\@secondoftwo}%
\providecommand \translation [1]{[#1]}%
\providecommand \BibitemOpen [0]{}%
\providecommand \bibitemStop [0]{}%
\providecommand \bibitemNoStop [0]{.\EOS\space}%
\providecommand \EOS [0]{\spacefactor3000\relax}%
\providecommand \BibitemShut  [1]{\csname bibitem#1\endcsname}%
\let\auto@bib@innerbib\@empty
\bibitem [{\citenamefont {Goold}\ \emph {et~al.}(2016)\citenamefont {Goold},
  \citenamefont {Huber}, \citenamefont {Riera}, \citenamefont {Rio},\ and\
  \citenamefont {Skrzypczyk}}]{Goold2016}%
  \BibitemOpen
  \bibfield  {author} {\bibinfo {author} {\bibfnamefont {J.}~\bibnamefont
  {Goold}}, \bibinfo {author} {\bibfnamefont {M.}~\bibnamefont {Huber}},
  \bibinfo {author} {\bibfnamefont {A.}~\bibnamefont {Riera}}, \bibinfo
  {author} {\bibfnamefont {L.~d.}\ \bibnamefont {Rio}}, \ and\ \bibinfo
  {author} {\bibfnamefont {P.}~\bibnamefont {Skrzypczyk}},\ }\href {\doibase
  10.1088/1751-8113/49/14/143001} {\bibfield  {journal} {\bibinfo  {journal}
  {J. Phys. A}\ }\textbf {\bibinfo {volume} {49}},\ \bibinfo {pages} {143001}
  (\bibinfo {year} {2016})}\BibitemShut {NoStop}%
\bibitem [{\citenamefont {Vinjanampathy}\ and\ \citenamefont
  {Anders}(2016)}]{Vinjanampathy2016}%
  \BibitemOpen
  \bibfield  {author} {\bibinfo {author} {\bibfnamefont {S.}~\bibnamefont
  {Vinjanampathy}}\ and\ \bibinfo {author} {\bibfnamefont {J.}~\bibnamefont
  {Anders}},\ }\href {\doibase 10.1080/00107514.2016.1201896} {\bibfield
  {journal} {\bibinfo  {journal} {Contemp. Phys.}\ }\textbf {\bibinfo {volume}
  {57}},\ \bibinfo {pages} {545–579} (\bibinfo {year} {2016})}\BibitemShut
  {NoStop}%
\bibitem [{\citenamefont {\AA{}berg}(2013)}]{aberg2013truly}%
  \BibitemOpen
  \bibfield  {author} {\bibinfo {author} {\bibfnamefont {J.}~\bibnamefont
  {\AA{}berg}},\ }\href {\doibase 10.1038/ncomms2712} {\bibfield  {journal}
  {\bibinfo  {journal} {Nat. Commun.}\ }\textbf {\bibinfo {volume} {4}},\
  \bibinfo {pages} {1925} (\bibinfo {year} {2013})}\BibitemShut {NoStop}%
\bibitem [{\citenamefont {Campisi}\ \emph
  {et~al.}(2011{\natexlab{a}})\citenamefont {Campisi}, \citenamefont
  {H\"anggi},\ and\ \citenamefont {Talkner}}]{TPM1RMP}%
  \BibitemOpen
  \bibfield  {author} {\bibinfo {author} {\bibfnamefont {M.}~\bibnamefont
  {Campisi}}, \bibinfo {author} {\bibfnamefont {P.}~\bibnamefont {H\"anggi}}, \
  and\ \bibinfo {author} {\bibfnamefont {P.}~\bibnamefont {Talkner}},\ }\href
  {\doibase 10.1103/RevModPhys.83.771} {\bibfield  {journal} {\bibinfo
  {journal} {Rev. Mod. Phys.}\ }\textbf {\bibinfo {volume} {83}},\ \bibinfo
  {pages} {771} (\bibinfo {year} {2011}{\natexlab{a}})}\BibitemShut {NoStop}%
\bibitem [{\citenamefont {Esposito}\ \emph {et~al.}(2009)\citenamefont
  {Esposito}, \citenamefont {Harbola},\ and\ \citenamefont
  {Mukamel}}]{Esposito2009nonequilibrium}%
  \BibitemOpen
  \bibfield  {author} {\bibinfo {author} {\bibfnamefont {M.}~\bibnamefont
  {Esposito}}, \bibinfo {author} {\bibfnamefont {U.}~\bibnamefont {Harbola}}, \
  and\ \bibinfo {author} {\bibfnamefont {S.}~\bibnamefont {Mukamel}},\ }\href
  {\doibase 10.1103/RevModPhys.81.1665} {\bibfield  {journal} {\bibinfo
  {journal} {Rev. Mod. Phys.}\ }\textbf {\bibinfo {volume} {81}},\ \bibinfo
  {pages} {1665} (\bibinfo {year} {2009})}\BibitemShut {NoStop}%
\bibitem [{\citenamefont {Esposito}\ \emph {et~al.}(2014)\citenamefont
  {Esposito}, \citenamefont {Harbola},\ and\ \citenamefont
  {Mukamel}}]{erratumm}%
  \BibitemOpen
  \bibfield  {author} {\bibinfo {author} {\bibfnamefont {M.}~\bibnamefont
  {Esposito}}, \bibinfo {author} {\bibfnamefont {U.}~\bibnamefont {Harbola}}, \
  and\ \bibinfo {author} {\bibfnamefont {S.}~\bibnamefont {Mukamel}},\ }\href
  {\doibase 10.1103/RevModPhys.86.1125} {\bibfield  {journal} {\bibinfo
  {journal} {Rev. Mod. Phys.}\ }\textbf {\bibinfo {volume} {86}},\ \bibinfo
  {pages} {1125} (\bibinfo {year} {2014})}\BibitemShut {NoStop}%
\bibitem [{\citenamefont {Jarzynski}(2011)}]{jarzynski2011equalities}%
  \BibitemOpen
  \bibfield  {author} {\bibinfo {author} {\bibfnamefont {C.}~\bibnamefont
  {Jarzynski}},\ }\href {\doibase 10.1146/annurev-conmatphys-062910-140506}
  {\bibfield  {journal} {\bibinfo  {journal} {Ann. Rev. Condens. Matter Phys.}\
  }\textbf {\bibinfo {volume} {2}},\ \bibinfo {pages} {329} (\bibinfo {year}
  {2011})}\BibitemShut {NoStop}%
\bibitem [{\citenamefont {Ro\ss{}nagel}\ \emph {et~al.}(2014)\citenamefont
  {Ro\ss{}nagel}, \citenamefont {Abah}, \citenamefont {Schmidt-Kaler},
  \citenamefont {Singer},\ and\ \citenamefont {Lutz}}]{Rossnagel2014nanoscale}%
  \BibitemOpen
  \bibfield  {author} {\bibinfo {author} {\bibfnamefont {J.}~\bibnamefont
  {Ro\ss{}nagel}}, \bibinfo {author} {\bibfnamefont {O.}~\bibnamefont {Abah}},
  \bibinfo {author} {\bibfnamefont {F.}~\bibnamefont {Schmidt-Kaler}}, \bibinfo
  {author} {\bibfnamefont {K.}~\bibnamefont {Singer}}, \ and\ \bibinfo {author}
  {\bibfnamefont {E.}~\bibnamefont {Lutz}},\ }\href {\doibase
  10.1103/PhysRevLett.112.030602} {\bibfield  {journal} {\bibinfo  {journal}
  {Phys. Rev. Lett.}\ }\textbf {\bibinfo {volume} {112}},\ \bibinfo {pages}
  {030602} (\bibinfo {year} {2014})}\BibitemShut {NoStop}%
\bibitem [{\citenamefont {Correa}\ \emph {et~al.}(2014)\citenamefont {Correa},
  \citenamefont {Palao}, \citenamefont {Alonso},\ and\ \citenamefont
  {Adesso}}]{Correa2014quantum}%
  \BibitemOpen
  \bibfield  {author} {\bibinfo {author} {\bibfnamefont {L.~A.}\ \bibnamefont
  {Correa}}, \bibinfo {author} {\bibfnamefont {J.~e.~P.}\ \bibnamefont
  {Palao}}, \bibinfo {author} {\bibfnamefont {D.}~\bibnamefont {Alonso}}, \
  and\ \bibinfo {author} {\bibfnamefont {G.}~\bibnamefont {Adesso}},\ }\href
  {\doibase 10.1038/srep03949} {\bibfield  {journal} {\bibinfo  {journal}
  {Scientific Reports}\ }\textbf {\bibinfo {volume} {4}},\ \bibinfo {pages}
  {3949} (\bibinfo {year} {2014})}\BibitemShut {NoStop}%
\bibitem [{\citenamefont {Alicki}\ and\ \citenamefont
  {Gelbwaser-Klimovsky}(2015)}]{Alicki2015Noneq}%
  \BibitemOpen
  \bibfield  {author} {\bibinfo {author} {\bibfnamefont {R.}~\bibnamefont
  {Alicki}}\ and\ \bibinfo {author} {\bibfnamefont {D.}~\bibnamefont
  {Gelbwaser-Klimovsky}},\ }\href
  {http://stacks.iop.org/1367-2630/17/i=11/a=115012} {\bibfield  {journal}
  {\bibinfo  {journal} {New Journal of Physics}\ }\textbf {\bibinfo {volume}
  {17}},\ \bibinfo {pages} {115012} (\bibinfo {year} {2015})}\BibitemShut
  {NoStop}%
\bibitem [{\citenamefont {Brask}\ and\ \citenamefont
  {Brunner}(2015)}]{Brask2015Small}%
  \BibitemOpen
  \bibfield  {author} {\bibinfo {author} {\bibfnamefont {J.~B.}\ \bibnamefont
  {Brask}}\ and\ \bibinfo {author} {\bibfnamefont {N.}~\bibnamefont
  {Brunner}},\ }\href {\doibase 10.1103/PhysRevE.92.062101} {\bibfield
  {journal} {\bibinfo  {journal} {Phys. Rev. E}\ }\textbf {\bibinfo {volume}
  {92}},\ \bibinfo {pages} {062101} (\bibinfo {year} {2015})}\BibitemShut
  {NoStop}%
\bibitem [{\citenamefont {{Uzdin}}\ \emph {et~al.}(2015)\citenamefont
  {{Uzdin}}, \citenamefont {{Levy}},\ and\ \citenamefont
  {{Kosloff}}}]{uzdin2015quantum}%
  \BibitemOpen
  \bibfield  {author} {\bibinfo {author} {\bibfnamefont {R.}~\bibnamefont
  {{Uzdin}}}, \bibinfo {author} {\bibfnamefont {A.}~\bibnamefont {{Levy}}}, \
  and\ \bibinfo {author} {\bibfnamefont {R.}~\bibnamefont {{Kosloff}}},\
  }\href@noop {} {\bibfield  {journal} {\bibinfo  {journal} {arXiv:1502.06592}\
  } (\bibinfo {year} {2015})}\BibitemShut {NoStop}%
\bibitem [{\citenamefont {Mitchison}\ \emph {et~al.}(2015)\citenamefont
  {Mitchison}, \citenamefont {Woods}, \citenamefont {Prior},\ and\
  \citenamefont {Huber}}]{Mitchison2015Coherence}%
  \BibitemOpen
  \bibfield  {author} {\bibinfo {author} {\bibfnamefont {M.~T.}\ \bibnamefont
  {Mitchison}}, \bibinfo {author} {\bibfnamefont {M.~P.}\ \bibnamefont
  {Woods}}, \bibinfo {author} {\bibfnamefont {J.}~\bibnamefont {Prior}}, \ and\
  \bibinfo {author} {\bibfnamefont {M.}~\bibnamefont {Huber}},\ }\href
  {http://stacks.iop.org/1367-2630/17/i=11/a=115013} {\bibfield  {journal}
  {\bibinfo  {journal} {New Journal of Physics}\ }\textbf {\bibinfo {volume}
  {17}},\ \bibinfo {pages} {115013} (\bibinfo {year} {2015})}\BibitemShut
  {NoStop}%
\bibitem [{\citenamefont {Hofer}\ \emph {et~al.}(2016)\citenamefont {Hofer},
  \citenamefont {Perarnau-Llobet}, \citenamefont {Brask}, \citenamefont
  {Silva}, \citenamefont {Huber},\ and\ \citenamefont
  {Brunner}}]{Hofer2016autonomous}%
  \BibitemOpen
  \bibfield  {author} {\bibinfo {author} {\bibfnamefont {P.~P.}\ \bibnamefont
  {Hofer}}, \bibinfo {author} {\bibfnamefont {M.}~\bibnamefont
  {Perarnau-Llobet}}, \bibinfo {author} {\bibfnamefont {J.~B.}\ \bibnamefont
  {Brask}}, \bibinfo {author} {\bibfnamefont {R.}~\bibnamefont {Silva}},
  \bibinfo {author} {\bibfnamefont {M.}~\bibnamefont {Huber}}, \ and\ \bibinfo
  {author} {\bibfnamefont {N.}~\bibnamefont {Brunner}},\ }\href {\doibase
  10.1103/PhysRevB.94.235420} {\bibfield  {journal} {\bibinfo  {journal} {Phys.
  Rev. B}\ }\textbf {\bibinfo {volume} {94}},\ \bibinfo {pages} {235420}
  (\bibinfo {year} {2016})}\BibitemShut {NoStop}%
\bibitem [{\citenamefont {Nimmrichter}\ \emph {et~al.}(2017)\citenamefont
  {Nimmrichter}, \citenamefont {Dai}, \citenamefont {Roulet},\ and\
  \citenamefont {Scarani}}]{Nimmrichter2017andclassical}%
  \BibitemOpen
  \bibfield  {author} {\bibinfo {author} {\bibfnamefont {S.}~\bibnamefont
  {Nimmrichter}}, \bibinfo {author} {\bibfnamefont {J.}~\bibnamefont {Dai}},
  \bibinfo {author} {\bibfnamefont {A.}~\bibnamefont {Roulet}}, \ and\ \bibinfo
  {author} {\bibfnamefont {V.}~\bibnamefont {Scarani}},\ }\href {\doibase
  10.22331/q-2017-12-11-37} {\bibfield  {journal} {\bibinfo  {journal}
  {{Quantum}}\ }\textbf {\bibinfo {volume} {1}},\ \bibinfo {pages} {37}
  (\bibinfo {year} {2017})}\BibitemShut {NoStop}%
\bibitem [{\citenamefont {Brandner}\ \emph {et~al.}(2017)\citenamefont
  {Brandner}, \citenamefont {Bauer},\ and\ \citenamefont
  {Seifert}}]{Brandner2017universal}%
  \BibitemOpen
  \bibfield  {author} {\bibinfo {author} {\bibfnamefont {K.}~\bibnamefont
  {Brandner}}, \bibinfo {author} {\bibfnamefont {M.}~\bibnamefont {Bauer}}, \
  and\ \bibinfo {author} {\bibfnamefont {U.}~\bibnamefont {Seifert}},\ }\href
  {\doibase 10.1103/PhysRevLett.119.170602} {\bibfield  {journal} {\bibinfo
  {journal} {Phys. Rev. Lett.}\ }\textbf {\bibinfo {volume} {119}},\ \bibinfo
  {pages} {170602} (\bibinfo {year} {2017})}\BibitemShut {NoStop}%
\bibitem [{\citenamefont {Klatzow}\ \emph {et~al.}(2017)\citenamefont
  {Klatzow}, \citenamefont {Weinzetl}, \citenamefont {Ledingham}, \citenamefont
  {Becker}, \citenamefont {Saunders}, \citenamefont {Nunn}, \citenamefont
  {Walmsley}, \citenamefont {Uzdin},\ and\ \citenamefont
  {Poem}}]{Klatzow2017Experimental}%
  \BibitemOpen
  \bibfield  {author} {\bibinfo {author} {\bibfnamefont {J.}~\bibnamefont
  {Klatzow}}, \bibinfo {author} {\bibfnamefont {C.}~\bibnamefont {Weinzetl}},
  \bibinfo {author} {\bibfnamefont {P.~M.}\ \bibnamefont {Ledingham}}, \bibinfo
  {author} {\bibfnamefont {J.~N.}\ \bibnamefont {Becker}}, \bibinfo {author}
  {\bibfnamefont {D.~J.}\ \bibnamefont {Saunders}}, \bibinfo {author}
  {\bibfnamefont {J.}~\bibnamefont {Nunn}}, \bibinfo {author} {\bibfnamefont
  {I.~A.}\ \bibnamefont {Walmsley}}, \bibinfo {author} {\bibfnamefont
  {R.}~\bibnamefont {Uzdin}}, \ and\ \bibinfo {author} {\bibfnamefont
  {E.}~\bibnamefont {Poem}},\ }\href {https://arxiv.org/abs/1710.08716}
  {\bibfield  {journal} {\bibinfo  {journal} {arXiv:1710.08716}\ } (\bibinfo
  {year} {2017})}\BibitemShut {NoStop}%
\bibitem [{\citenamefont {Allahverdyan}\ \emph {et~al.}(2004)\citenamefont
  {Allahverdyan}, \citenamefont {Balian},\ and\ \citenamefont
  {Nieuwenhuizen}}]{Allahverdyan2004maximal}%
  \BibitemOpen
  \bibfield  {author} {\bibinfo {author} {\bibfnamefont {A.~E.}\ \bibnamefont
  {Allahverdyan}}, \bibinfo {author} {\bibfnamefont {R.}~\bibnamefont
  {Balian}}, \ and\ \bibinfo {author} {\bibfnamefont {T.~M.}\ \bibnamefont
  {Nieuwenhuizen}},\ }\href {http://stacks.iop.org/0295-5075/67/i=4/a=565}
  {\bibfield  {journal} {\bibinfo  {journal} {EPL (Europhysics Letters)}\
  }\textbf {\bibinfo {volume} {67}},\ \bibinfo {pages} {565} (\bibinfo {year}
  {2004})}\BibitemShut {NoStop}%
\bibitem [{\citenamefont {Funo}\ \emph {et~al.}(2013)\citenamefont {Funo},
  \citenamefont {Watanabe},\ and\ \citenamefont {Ueda}}]{Funo2013work}%
  \BibitemOpen
  \bibfield  {author} {\bibinfo {author} {\bibfnamefont {K.}~\bibnamefont
  {Funo}}, \bibinfo {author} {\bibfnamefont {Y.}~\bibnamefont {Watanabe}}, \
  and\ \bibinfo {author} {\bibfnamefont {M.}~\bibnamefont {Ueda}},\ }\href
  {\doibase 10.1103/PhysRevA.88.052319} {\bibfield  {journal} {\bibinfo
  {journal} {Phys. Rev. A}\ }\textbf {\bibinfo {volume} {88}},\ \bibinfo
  {pages} {052319} (\bibinfo {year} {2013})}\BibitemShut {NoStop}%
\bibitem [{\citenamefont {Perarnau-Llobet}\ \emph {et~al.}(2015)\citenamefont
  {Perarnau-Llobet}, \citenamefont {Hovhannisyan}, \citenamefont {Huber},
  \citenamefont {Skrzypczyk}, \citenamefont {Brunner},\ and\ \citenamefont
  {Ac\'{\i}n}}]{mpl2015extractable}%
  \BibitemOpen
  \bibfield  {author} {\bibinfo {author} {\bibfnamefont {M.}~\bibnamefont
  {Perarnau-Llobet}}, \bibinfo {author} {\bibfnamefont {K.~V.}\ \bibnamefont
  {Hovhannisyan}}, \bibinfo {author} {\bibfnamefont {M.}~\bibnamefont {Huber}},
  \bibinfo {author} {\bibfnamefont {P.}~\bibnamefont {Skrzypczyk}}, \bibinfo
  {author} {\bibfnamefont {N.}~\bibnamefont {Brunner}}, \ and\ \bibinfo
  {author} {\bibfnamefont {A.}~\bibnamefont {Ac\'{\i}n}},\ }\href {\doibase
  10.1103/PhysRevX.5.041011} {\bibfield  {journal} {\bibinfo  {journal} {Phys.
  Rev. X}\ }\textbf {\bibinfo {volume} {5}},\ \bibinfo {pages} {041011}
  (\bibinfo {year} {2015})}\BibitemShut {NoStop}%
\bibitem [{\citenamefont {Korzekwa}\ \emph {et~al.}(2016)\citenamefont
  {Korzekwa}, \citenamefont {Lostaglio}, \citenamefont {Oppenheim},\ and\
  \citenamefont {Jennings}}]{korzekwa2016extraction}%
  \BibitemOpen
  \bibfield  {author} {\bibinfo {author} {\bibfnamefont {K.}~\bibnamefont
  {Korzekwa}}, \bibinfo {author} {\bibfnamefont {M.}~\bibnamefont {Lostaglio}},
  \bibinfo {author} {\bibfnamefont {J.}~\bibnamefont {Oppenheim}}, \ and\
  \bibinfo {author} {\bibfnamefont {D.}~\bibnamefont {Jennings}},\ }\href
  {http://iopscience.iop.org/article/10.1088/1367-2630/18/2/023045/meta}
  {\bibfield  {journal} {\bibinfo  {journal} {New Journal of Physics}\ }\textbf
  {\bibinfo {volume} {18}},\ \bibinfo {pages} {023045} (\bibinfo {year}
  {2016})}\BibitemShut {NoStop}%
\bibitem [{\citenamefont {Misra}\ \emph {et~al.}(2016)\citenamefont {Misra},
  \citenamefont {Singh}, \citenamefont {Bhattacharya},\ and\ \citenamefont
  {Pati}}]{EnergycostPRA2016}%
  \BibitemOpen
  \bibfield  {author} {\bibinfo {author} {\bibfnamefont {A.}~\bibnamefont
  {Misra}}, \bibinfo {author} {\bibfnamefont {U.}~\bibnamefont {Singh}},
  \bibinfo {author} {\bibfnamefont {S.}~\bibnamefont {Bhattacharya}}, \ and\
  \bibinfo {author} {\bibfnamefont {A.~K.}\ \bibnamefont {Pati}},\ }\href
  {\doibase 10.1103/PhysRevA.93.052335} {\bibfield  {journal} {\bibinfo
  {journal} {Phys. Rev. A}\ }\textbf {\bibinfo {volume} {93}},\ \bibinfo
  {pages} {052335} (\bibinfo {year} {2016})}\BibitemShut {NoStop}%
\bibitem [{\citenamefont {L\"orch}\ \emph {et~al.}(2018)\citenamefont
  {L\"orch}, \citenamefont {Bruder}, \citenamefont {Brunner},\ and\
  \citenamefont {P.~Hofer}}]{Patrick2018}%
  \BibitemOpen
  \bibfield  {author} {\bibinfo {author} {\bibfnamefont {N.}~\bibnamefont
  {L\"orch}}, \bibinfo {author} {\bibfnamefont {C.}~\bibnamefont {Bruder}},
  \bibinfo {author} {\bibfnamefont {N.}~\bibnamefont {Brunner}}, \ and\
  \bibinfo {author} {\bibfnamefont {P.}~\bibnamefont {P.~Hofer}},\ }\href
  {https://arxiv.org/abs/1802.10572} {\bibfield  {journal} {\bibinfo  {journal}
  {arXiv:1802.10572}\ } (\bibinfo {year} {2018})}\BibitemShut {NoStop}%
\bibitem [{\citenamefont {Hovhannisyan}\ \emph {et~al.}(2013)\citenamefont
  {Hovhannisyan}, \citenamefont {Perarnau-Llobet}, \citenamefont {Huber},\ and\
  \citenamefont {Ac\'{\i}n}}]{Karen2013}%
  \BibitemOpen
  \bibfield  {author} {\bibinfo {author} {\bibfnamefont {K.~V.}\ \bibnamefont
  {Hovhannisyan}}, \bibinfo {author} {\bibfnamefont {M.}~\bibnamefont
  {Perarnau-Llobet}}, \bibinfo {author} {\bibfnamefont {M.}~\bibnamefont
  {Huber}}, \ and\ \bibinfo {author} {\bibfnamefont {A.}~\bibnamefont
  {Ac\'{\i}n}},\ }\href {\doibase 10.1103/PhysRevLett.111.240401} {\bibfield
  {journal} {\bibinfo  {journal} {Phys. Rev. Lett.}\ }\textbf {\bibinfo
  {volume} {111}},\ \bibinfo {pages} {240401} (\bibinfo {year}
  {2013})}\BibitemShut {NoStop}%
\bibitem [{\citenamefont {Brunner}\ \emph
  {et~al.}(2014{\natexlab{a}})\citenamefont {Brunner}, \citenamefont {Huber},
  \citenamefont {Linden}, \citenamefont {Popescu}, \citenamefont {Silva},\ and\
  \citenamefont {Skrzypczyk}}]{Brunner2014entanglement}%
  \BibitemOpen
  \bibfield  {author} {\bibinfo {author} {\bibfnamefont {N.}~\bibnamefont
  {Brunner}}, \bibinfo {author} {\bibfnamefont {M.}~\bibnamefont {Huber}},
  \bibinfo {author} {\bibfnamefont {N.}~\bibnamefont {Linden}}, \bibinfo
  {author} {\bibfnamefont {S.}~\bibnamefont {Popescu}}, \bibinfo {author}
  {\bibfnamefont {R.}~\bibnamefont {Silva}}, \ and\ \bibinfo {author}
  {\bibfnamefont {P.}~\bibnamefont {Skrzypczyk}},\ }\href {\doibase
  10.1103/PhysRevE.89.032115} {\bibfield  {journal} {\bibinfo  {journal} {Phys.
  Rev. E}\ }\textbf {\bibinfo {volume} {89}},\ \bibinfo {pages} {032115}
  (\bibinfo {year} {2014}{\natexlab{a}})}\BibitemShut {NoStop}%
\bibitem [{\citenamefont {Uzdin}\ \emph {et~al.}(2015)\citenamefont {Uzdin},
  \citenamefont {Levy},\ and\ \citenamefont {Kosloff}}]{Raam2015equivalence}%
  \BibitemOpen
  \bibfield  {author} {\bibinfo {author} {\bibfnamefont {R.}~\bibnamefont
  {Uzdin}}, \bibinfo {author} {\bibfnamefont {A.}~\bibnamefont {Levy}}, \ and\
  \bibinfo {author} {\bibfnamefont {R.}~\bibnamefont {Kosloff}},\ }\href
  {\doibase 10.1103/PhysRevX.5.031044} {\bibfield  {journal} {\bibinfo
  {journal} {Phys. Rev. X}\ }\textbf {\bibinfo {volume} {5}},\ \bibinfo {pages}
  {031044} (\bibinfo {year} {2015})}\BibitemShut {NoStop}%
\bibitem [{\citenamefont {Campaioli}\ \emph {et~al.}(2017)\citenamefont
  {Campaioli}, \citenamefont {Pollock}, \citenamefont {Binder}, \citenamefont
  {C\'eleri}, \citenamefont {Goold}, \citenamefont {Vinjanampathy},\ and\
  \citenamefont {Modi}}]{Campaioli2017}%
  \BibitemOpen
  \bibfield  {author} {\bibinfo {author} {\bibfnamefont {F.}~\bibnamefont
  {Campaioli}}, \bibinfo {author} {\bibfnamefont {F.~A.}\ \bibnamefont
  {Pollock}}, \bibinfo {author} {\bibfnamefont {F.~C.}\ \bibnamefont {Binder}},
  \bibinfo {author} {\bibfnamefont {L.}~\bibnamefont {C\'eleri}}, \bibinfo
  {author} {\bibfnamefont {J.}~\bibnamefont {Goold}}, \bibinfo {author}
  {\bibfnamefont {S.}~\bibnamefont {Vinjanampathy}}, \ and\ \bibinfo {author}
  {\bibfnamefont {K.}~\bibnamefont {Modi}},\ }\href {\doibase
  10.1103/PhysRevLett.118.150601} {\bibfield  {journal} {\bibinfo  {journal}
  {Phys. Rev. Lett.}\ }\textbf {\bibinfo {volume} {118}},\ \bibinfo {pages}
  {150601} (\bibinfo {year} {2017})}\BibitemShut {NoStop}%
\bibitem [{\citenamefont {Ferraro}\ \emph {et~al.}(2018)\citenamefont
  {Ferraro}, \citenamefont {Campisi}, \citenamefont {Andolina}, \citenamefont
  {Pellegrini},\ and\ \citenamefont {Polini}}]{Ferraro2017}%
  \BibitemOpen
  \bibfield  {author} {\bibinfo {author} {\bibfnamefont {D.}~\bibnamefont
  {Ferraro}}, \bibinfo {author} {\bibfnamefont {M.}~\bibnamefont {Campisi}},
  \bibinfo {author} {\bibfnamefont {G.~M.}\ \bibnamefont {Andolina}}, \bibinfo
  {author} {\bibfnamefont {V.}~\bibnamefont {Pellegrini}}, \ and\ \bibinfo
  {author} {\bibfnamefont {M.}~\bibnamefont {Polini}},\ }\href {\doibase
  10.1103/PhysRevLett.120.117702} {\bibfield  {journal} {\bibinfo  {journal}
  {Phys. Rev. Lett.}\ }\textbf {\bibinfo {volume} {120}},\ \bibinfo {pages}
  {117702} (\bibinfo {year} {2018})}\BibitemShut {NoStop}%
\bibitem [{\citenamefont {Watanabe}\ \emph {et~al.}(2017)\citenamefont
  {Watanabe}, \citenamefont {Venkatesh}, \citenamefont {Talkner},\ and\
  \citenamefont {del Campo}}]{watanabe2017quantum}%
  \BibitemOpen
  \bibfield  {author} {\bibinfo {author} {\bibfnamefont {G.}~\bibnamefont
  {Watanabe}}, \bibinfo {author} {\bibfnamefont {B.~P.}\ \bibnamefont
  {Venkatesh}}, \bibinfo {author} {\bibfnamefont {P.}~\bibnamefont {Talkner}},
  \ and\ \bibinfo {author} {\bibfnamefont {A.}~\bibnamefont {del Campo}},\
  }\href {\doibase 10.1103/PhysRevLett.118.050601} {\bibfield  {journal}
  {\bibinfo  {journal} {Phys. Rev. Lett.}\ }\textbf {\bibinfo {volume} {118}},\
  \bibinfo {pages} {050601} (\bibinfo {year} {2017})}\BibitemShut {NoStop}%
\bibitem [{\citenamefont {Levy}\ and\ \citenamefont
  {Gelbwaser-Klimovsky}(2018{\natexlab{a}})}]{levy2018quantum}%
  \BibitemOpen
  \bibfield  {author} {\bibinfo {author} {\bibfnamefont {A.}~\bibnamefont
  {Levy}}\ and\ \bibinfo {author} {\bibfnamefont {D.}~\bibnamefont
  {Gelbwaser-Klimovsky}},\ }\href {https://arxiv.org/abs/1803.05586} {\bibfield
   {journal} {\bibinfo  {journal} {arXiv:1803.05586}\ } (\bibinfo {year}
  {2018}{\natexlab{a}})}\BibitemShut {NoStop}%
\bibitem [{\citenamefont {Busch}\ \emph {et~al.}(2014)\citenamefont {Busch},
  \citenamefont {Lahti},\ and\ \citenamefont {Werner}}]{busch2014quantum}%
  \BibitemOpen
  \bibfield  {author} {\bibinfo {author} {\bibfnamefont {P.}~\bibnamefont
  {Busch}}, \bibinfo {author} {\bibfnamefont {P.}~\bibnamefont {Lahti}}, \ and\
  \bibinfo {author} {\bibfnamefont {R.~F.}\ \bibnamefont {Werner}},\ }\href
  {\doibase 10.1103/RevModPhys.86.1261} {\bibfield  {journal} {\bibinfo
  {journal} {Rev. Mod. Phys.}\ }\textbf {\bibinfo {volume} {86}},\ \bibinfo
  {pages} {1261} (\bibinfo {year} {2014})}\BibitemShut {NoStop}%
\bibitem [{\citenamefont {Jarzynski}(2008)}]{Jarzynskia2008}%
  \BibitemOpen
  \bibfield  {author} {\bibinfo {author} {\bibfnamefont {C.}~\bibnamefont
  {Jarzynski}},\ }\href {\doibase 10.1140/epjb/e2008-00254-2} {\bibfield
  {journal} {\bibinfo  {journal} {The European Physical Journal B}\ }\textbf
  {\bibinfo {volume} {64}},\ \bibinfo {pages} {331} (\bibinfo {year}
  {2008})}\BibitemShut {NoStop}%
\bibitem [{\citenamefont {{Bochkov}}\ and\ \citenamefont
  {{Kuzovlev}}(1977)}]{Bochkov1977}%
  \BibitemOpen
  \bibfield  {author} {\bibinfo {author} {\bibfnamefont {G.~N.}\ \bibnamefont
  {{Bochkov}}}\ and\ \bibinfo {author} {\bibfnamefont {I.~E.}\ \bibnamefont
  {{Kuzovlev}}},\ }\href@noop {} {\bibfield  {journal} {\bibinfo  {journal}
  {Zhurnal Eksperimentalnoi i Teoreticheskoi Fiziki}\ }\textbf {\bibinfo
  {volume} {72}},\ \bibinfo {pages} {238} (\bibinfo {year} {1977})}\BibitemShut
  {NoStop}%
\bibitem [{\citenamefont {Bochkov}\ and\ \citenamefont
  {Kuzovlev}(1981)}]{BOCHKOV1981443}%
  \BibitemOpen
  \bibfield  {author} {\bibinfo {author} {\bibfnamefont {G.}~\bibnamefont
  {Bochkov}}\ and\ \bibinfo {author} {\bibfnamefont {Y.}~\bibnamefont
  {Kuzovlev}},\ }\href {\doibase https://doi.org/10.1016/0378-4371(81)90122-9}
  {\bibfield  {journal} {\bibinfo  {journal} {Physica A: Statistical Mechanics
  and its Applications}\ }\textbf {\bibinfo {volume} {106}},\ \bibinfo {pages}
  {443 } (\bibinfo {year} {1981})}\BibitemShut {NoStop}%
\bibitem [{\citenamefont {Yukawa}(2000)}]{Yukawa2000}%
  \BibitemOpen
  \bibfield  {author} {\bibinfo {author} {\bibfnamefont {S.}~\bibnamefont
  {Yukawa}},\ }\href {\doibase 10.1143/JPSJ.69.2367} {\bibfield  {journal}
  {\bibinfo  {journal} {Journal of the Physical Society of Japan}\ }\textbf
  {\bibinfo {volume} {69}},\ \bibinfo {pages} {2367} (\bibinfo {year}
  {2000})},\ \Eprint
  {http://arxiv.org/abs/https://doi.org/10.1143/JPSJ.69.2367}
  {https://doi.org/10.1143/JPSJ.69.2367} \BibitemShut {NoStop}%
\bibitem [{\citenamefont {Monnai}\ and\ \citenamefont
  {Tasaki}(2003)}]{Monnai2003}%
  \BibitemOpen
  \bibfield  {author} {\bibinfo {author} {\bibfnamefont {T.}~\bibnamefont
  {Monnai}}\ and\ \bibinfo {author} {\bibfnamefont {S.}~\bibnamefont
  {Tasaki}},\ }\href@noop {} {\bibfield  {journal} {\bibinfo  {journal}
  {arXiv:1707.04930}\ } (\bibinfo {year} {2003})}\BibitemShut {NoStop}%
\bibitem [{\citenamefont {Chernyak}\ and\ \citenamefont
  {Mukamel}(2004)}]{Chernyak2004}%
  \BibitemOpen
  \bibfield  {author} {\bibinfo {author} {\bibfnamefont {V.}~\bibnamefont
  {Chernyak}}\ and\ \bibinfo {author} {\bibfnamefont {S.}~\bibnamefont
  {Mukamel}},\ }\href {\doibase 10.1103/PhysRevLett.93.048302} {\bibfield
  {journal} {\bibinfo  {journal} {Phys. Rev. Lett.}\ }\textbf {\bibinfo
  {volume} {93}},\ \bibinfo {pages} {048302} (\bibinfo {year}
  {2004})}\BibitemShut {NoStop}%
\bibitem [{\citenamefont {Allahverdyan}\ and\ \citenamefont
  {Nieuwenhuizen}(2005)}]{Allahverdyan2005}%
  \BibitemOpen
  \bibfield  {author} {\bibinfo {author} {\bibfnamefont {A.~E.}\ \bibnamefont
  {Allahverdyan}}\ and\ \bibinfo {author} {\bibfnamefont {T.~M.}\ \bibnamefont
  {Nieuwenhuizen}},\ }\href {\doibase 10.1103/PhysRevE.71.066102} {\bibfield
  {journal} {\bibinfo  {journal} {Phys. Rev. E}\ }\textbf {\bibinfo {volume}
  {71}},\ \bibinfo {pages} {066102} (\bibinfo {year} {2005})}\BibitemShut
  {NoStop}%
\bibitem [{\citenamefont {Engel}\ and\ \citenamefont
  {Nolte}(2007)}]{Engel2007}%
  \BibitemOpen
  \bibfield  {author} {\bibinfo {author} {\bibfnamefont {A.}~\bibnamefont
  {Engel}}\ and\ \bibinfo {author} {\bibfnamefont {R.}~\bibnamefont {Nolte}},\
  }\href {http://stacks.iop.org/0295-5075/79/i=1/a=10003} {\bibfield  {journal}
  {\bibinfo  {journal} {EPL (Europhysics Letters)}\ }\textbf {\bibinfo {volume}
  {79}},\ \bibinfo {pages} {10003} (\bibinfo {year} {2007})}\BibitemShut
  {NoStop}%
\bibitem [{\citenamefont {Gelin}\ and\ \citenamefont
  {Kosov}(2008)}]{Gelin2008}%
  \BibitemOpen
  \bibfield  {author} {\bibinfo {author} {\bibfnamefont {M.~F.}\ \bibnamefont
  {Gelin}}\ and\ \bibinfo {author} {\bibfnamefont {D.~S.}\ \bibnamefont
  {Kosov}},\ }\href {\doibase 10.1103/PhysRevE.78.011116} {\bibfield  {journal}
  {\bibinfo  {journal} {Phys. Rev. E}\ }\textbf {\bibinfo {volume} {78}},\
  \bibinfo {pages} {011116} (\bibinfo {year} {2008})}\BibitemShut {NoStop}%
\bibitem [{\citenamefont {Kurchan}(2000)}]{TPM3arxiv}%
  \BibitemOpen
  \bibfield  {author} {\bibinfo {author} {\bibfnamefont {J.}~\bibnamefont
  {Kurchan}},\ }\href {https://arxiv.org/abs/cond-mat/0007360} {\bibfield
  {journal} {\bibinfo  {journal} {arXiv preprint cond-mat/0007360}\ } (\bibinfo
  {year} {2000})}\BibitemShut {NoStop}%
\bibitem [{\citenamefont {Tasaki}(2000)}]{TPM4arxiv}%
  \BibitemOpen
  \bibfield  {author} {\bibinfo {author} {\bibfnamefont {H.}~\bibnamefont
  {Tasaki}},\ }\href {https://arxiv.org/abs/cond-mat/0009244} {\bibfield
  {journal} {\bibinfo  {journal} {arXiv preprint cond-mat/0009244}\ } (\bibinfo
  {year} {2000})}\BibitemShut {NoStop}%
\bibitem [{\citenamefont {Talkner}\ \emph {et~al.}(2007)\citenamefont
  {Talkner}, \citenamefont {Lutz},\ and\ \citenamefont {H\"anggi}}]{TPM2PRE}%
  \BibitemOpen
  \bibfield  {author} {\bibinfo {author} {\bibfnamefont {P.}~\bibnamefont
  {Talkner}}, \bibinfo {author} {\bibfnamefont {E.}~\bibnamefont {Lutz}}, \
  and\ \bibinfo {author} {\bibfnamefont {P.}~\bibnamefont {H\"anggi}},\ }\href
  {\doibase 10.1103/PhysRevE.75.050102} {\bibfield  {journal} {\bibinfo
  {journal} {Phys. Rev. E}\ }\textbf {\bibinfo {volume} {75}},\ \bibinfo
  {pages} {050102} (\bibinfo {year} {2007})}\BibitemShut {NoStop}%
\bibitem [{\citenamefont {H\"anggi}\ and\ \citenamefont
  {Talkner}(2017)}]{Hanggi2015theother}%
  \BibitemOpen
  \bibfield  {author} {\bibinfo {author} {\bibfnamefont {P.}~\bibnamefont
  {H\"anggi}}\ and\ \bibinfo {author} {\bibfnamefont {P.}~\bibnamefont
  {Talkner}},\ }\href {\doibase 10.1038/nphys3167} {\bibfield  {journal}
  {\bibinfo  {journal} {Nature Physics}\ }\textbf {\bibinfo {volume} {11}},\
  \bibinfo {pages} {108} (\bibinfo {year} {2017})}\BibitemShut {NoStop}%
\bibitem [{\citenamefont {Campisi}\ \emph
  {et~al.}(2011{\natexlab{b}})\citenamefont {Campisi}, \citenamefont
  {H\"anggi},\ and\ \citenamefont {Talkner}}]{Campisi2011Colloquim}%
  \BibitemOpen
  \bibfield  {author} {\bibinfo {author} {\bibfnamefont {M.}~\bibnamefont
  {Campisi}}, \bibinfo {author} {\bibfnamefont {P.}~\bibnamefont {H\"anggi}}, \
  and\ \bibinfo {author} {\bibfnamefont {P.}~\bibnamefont {Talkner}},\ }\href
  {\doibase 10.1103/RevModPhys.83.771} {\bibfield  {journal} {\bibinfo
  {journal} {Rev. Mod. Phys.}\ }\textbf {\bibinfo {volume} {83}},\ \bibinfo
  {pages} {771} (\bibinfo {year} {2011}{\natexlab{b}})}\BibitemShut {NoStop}%
\bibitem [{\citenamefont {Campisi}\ \emph
  {et~al.}(2011{\natexlab{c}})\citenamefont {Campisi}, \citenamefont
  {H\"anggi},\ and\ \citenamefont {Talkner}}]{Campisi2011ColloquimII}%
  \BibitemOpen
  \bibfield  {author} {\bibinfo {author} {\bibfnamefont {M.}~\bibnamefont
  {Campisi}}, \bibinfo {author} {\bibfnamefont {P.}~\bibnamefont {H\"anggi}}, \
  and\ \bibinfo {author} {\bibfnamefont {P.}~\bibnamefont {Talkner}},\ }\href
  {\doibase 10.1103/RevModPhys.83.1653} {\bibfield  {journal} {\bibinfo
  {journal} {Rev. Mod. Phys.}\ }\textbf {\bibinfo {volume} {83}},\ \bibinfo
  {pages} {1653} (\bibinfo {year} {2011}{\natexlab{c}})}\BibitemShut {NoStop}%
\bibitem [{\citenamefont {Campisi}\ \emph {et~al.}(2009)\citenamefont
  {Campisi}, \citenamefont {Talkner},\ and\ \citenamefont
  {H\"anggi}}]{Campisi2009}%
  \BibitemOpen
  \bibfield  {author} {\bibinfo {author} {\bibfnamefont {M.}~\bibnamefont
  {Campisi}}, \bibinfo {author} {\bibfnamefont {P.}~\bibnamefont {Talkner}}, \
  and\ \bibinfo {author} {\bibfnamefont {P.}~\bibnamefont {H\"anggi}},\ }\href
  {\doibase 10.1103/PhysRevLett.102.210401} {\bibfield  {journal} {\bibinfo
  {journal} {Phys. Rev. Lett.}\ }\textbf {\bibinfo {volume} {102}},\ \bibinfo
  {pages} {210401} (\bibinfo {year} {2009})}\BibitemShut {NoStop}%
\bibitem [{\citenamefont {Huber}\ \emph {et~al.}(2008)\citenamefont {Huber},
  \citenamefont {Schmidt-Kaler}, \citenamefont {Deffner},\ and\ \citenamefont
  {Lutz}}]{Huber2008employing}%
  \BibitemOpen
  \bibfield  {author} {\bibinfo {author} {\bibfnamefont {G.}~\bibnamefont
  {Huber}}, \bibinfo {author} {\bibfnamefont {F.}~\bibnamefont
  {Schmidt-Kaler}}, \bibinfo {author} {\bibfnamefont {S.}~\bibnamefont
  {Deffner}}, \ and\ \bibinfo {author} {\bibfnamefont {E.}~\bibnamefont
  {Lutz}},\ }\href {\doibase 10.1103/PhysRevLett.101.070403} {\bibfield
  {journal} {\bibinfo  {journal} {Phys. Rev. Lett.}\ }\textbf {\bibinfo
  {volume} {101}},\ \bibinfo {pages} {070403} (\bibinfo {year}
  {2008})}\BibitemShut {NoStop}%
\bibitem [{\citenamefont {Batalh\~ao}\ \emph {et~al.}(2014)\citenamefont
  {Batalh\~ao}, \citenamefont {Souza}, \citenamefont {Mazzola}, \citenamefont
  {Auccaise}, \citenamefont {Sarthour}, \citenamefont {Oliveira}, \citenamefont
  {Goold}, \citenamefont {De~Chiara}, \citenamefont {Paternostro},\ and\
  \citenamefont {Serra}}]{Tiago2014experimental}%
  \BibitemOpen
  \bibfield  {author} {\bibinfo {author} {\bibfnamefont {T.~B.}\ \bibnamefont
  {Batalh\~ao}}, \bibinfo {author} {\bibfnamefont {A.~M.}\ \bibnamefont
  {Souza}}, \bibinfo {author} {\bibfnamefont {L.}~\bibnamefont {Mazzola}},
  \bibinfo {author} {\bibfnamefont {R.}~\bibnamefont {Auccaise}}, \bibinfo
  {author} {\bibfnamefont {R.~S.}\ \bibnamefont {Sarthour}}, \bibinfo {author}
  {\bibfnamefont {I.~S.}\ \bibnamefont {Oliveira}}, \bibinfo {author}
  {\bibfnamefont {J.}~\bibnamefont {Goold}}, \bibinfo {author} {\bibfnamefont
  {G.}~\bibnamefont {De~Chiara}}, \bibinfo {author} {\bibfnamefont
  {M.}~\bibnamefont {Paternostro}}, \ and\ \bibinfo {author} {\bibfnamefont
  {R.~M.}\ \bibnamefont {Serra}},\ }\href {\doibase
  10.1103/PhysRevLett.113.140601} {\bibfield  {journal} {\bibinfo  {journal}
  {Phys. Rev. Lett.}\ }\textbf {\bibinfo {volume} {113}},\ \bibinfo {pages}
  {140601} (\bibinfo {year} {2014})}\BibitemShut {NoStop}%
\bibitem [{\citenamefont {An}\ \emph {et~al.}(2014)\citenamefont {An},
  \citenamefont {Zhang}, \citenamefont {Um}, \citenamefont {Lv}, \citenamefont
  {Lu}, \citenamefont {Zhang}, \citenamefont {Yin}, \citenamefont {Quan},\ and\
  \citenamefont {Kim}}]{Shuoming2014experimental}%
  \BibitemOpen
  \bibfield  {author} {\bibinfo {author} {\bibfnamefont {S.}~\bibnamefont
  {An}}, \bibinfo {author} {\bibfnamefont {J.-N.}\ \bibnamefont {Zhang}},
  \bibinfo {author} {\bibfnamefont {M.}~\bibnamefont {Um}}, \bibinfo {author}
  {\bibfnamefont {D.}~\bibnamefont {Lv}}, \bibinfo {author} {\bibfnamefont
  {Y.}~\bibnamefont {Lu}}, \bibinfo {author} {\bibfnamefont {J.}~\bibnamefont
  {Zhang}}, \bibinfo {author} {\bibfnamefont {Z.-Q.}\ \bibnamefont {Yin}},
  \bibinfo {author} {\bibfnamefont {H.~T.}\ \bibnamefont {Quan}}, \ and\
  \bibinfo {author} {\bibfnamefont {K.}~\bibnamefont {Kim}},\ }\href {\doibase
  10.1038/nphys3197} {\bibfield  {journal} {\bibinfo  {journal} {Nature
  Physics}\ }\textbf {\bibinfo {volume} {11}},\ \bibinfo {pages} {193}
  (\bibinfo {year} {2014})}\BibitemShut {NoStop}%
\bibitem [{\citenamefont {Cerisola}\ \emph {et~al.}(2017)\citenamefont
  {Cerisola}, \citenamefont {Margalit}, \citenamefont {Machluf}, \citenamefont
  {Roncaglia}, \citenamefont {Paz},\ and\ \citenamefont
  {Folman}}]{Cerisola2017quantum}%
  \BibitemOpen
  \bibfield  {author} {\bibinfo {author} {\bibfnamefont {F.}~\bibnamefont
  {Cerisola}}, \bibinfo {author} {\bibfnamefont {Y.}~\bibnamefont {Margalit}},
  \bibinfo {author} {\bibfnamefont {S.}~\bibnamefont {Machluf}}, \bibinfo
  {author} {\bibfnamefont {A.~J.}\ \bibnamefont {Roncaglia}}, \bibinfo {author}
  {\bibfnamefont {J.~P.}\ \bibnamefont {Paz}}, \ and\ \bibinfo {author}
  {\bibfnamefont {R.}~\bibnamefont {Folman}},\ }\href {\doibase
  10.1038/s41467-017-01308-7} {\bibfield  {journal} {\bibinfo  {journal}
  {Nature Communications}\ }\textbf {\bibinfo {volume} {8}},\ \bibinfo {pages}
  {1241} (\bibinfo {year} {2017})}\BibitemShut {NoStop}%
\bibitem [{\citenamefont {Jarzynski}\ \emph {et~al.}(2015)\citenamefont
  {Jarzynski}, \citenamefont {Quan},\ and\ \citenamefont
  {Rahav}}]{Jarzynski2015quantum}%
  \BibitemOpen
  \bibfield  {author} {\bibinfo {author} {\bibfnamefont {C.}~\bibnamefont
  {Jarzynski}}, \bibinfo {author} {\bibfnamefont {H.~T.}\ \bibnamefont {Quan}},
  \ and\ \bibinfo {author} {\bibfnamefont {S.}~\bibnamefont {Rahav}},\ }\href
  {\doibase 10.1103/PhysRevX.5.031038} {\bibfield  {journal} {\bibinfo
  {journal} {Phys. Rev. X}\ }\textbf {\bibinfo {volume} {5}},\ \bibinfo {pages}
  {031038} (\bibinfo {year} {2015})}\BibitemShut {NoStop}%
\bibitem [{\citenamefont {Zhu}\ \emph {et~al.}(2016)\citenamefont {Zhu},
  \citenamefont {Gong}, \citenamefont {Wu},\ and\ \citenamefont
  {Quan}}]{Zhu2016quantum}%
  \BibitemOpen
  \bibfield  {author} {\bibinfo {author} {\bibfnamefont {L.}~\bibnamefont
  {Zhu}}, \bibinfo {author} {\bibfnamefont {Z.}~\bibnamefont {Gong}}, \bibinfo
  {author} {\bibfnamefont {B.}~\bibnamefont {Wu}}, \ and\ \bibinfo {author}
  {\bibfnamefont {H.~T.}\ \bibnamefont {Quan}},\ }\href {\doibase
  10.1103/PhysRevE.93.062108} {\bibfield  {journal} {\bibinfo  {journal} {Phys.
  Rev. E}\ }\textbf {\bibinfo {volume} {93}},\ \bibinfo {pages} {062108}
  (\bibinfo {year} {2016})}\BibitemShut {NoStop}%
\bibitem [{\citenamefont {Allahverdyan}(2014{\natexlab{a}})}]{NQFoWPRE2014}%
  \BibitemOpen
  \bibfield  {author} {\bibinfo {author} {\bibfnamefont {A.~E.}\ \bibnamefont
  {Allahverdyan}},\ }\href {\doibase 10.1103/PhysRevE.90.032137} {\bibfield
  {journal} {\bibinfo  {journal} {Phys. Rev. E}\ }\textbf {\bibinfo {volume}
  {90}},\ \bibinfo {pages} {032137} (\bibinfo {year}
  {2014}{\natexlab{a}})}\BibitemShut {NoStop}%
\bibitem [{\citenamefont {Solinas}\ and\ \citenamefont
  {Gasparinetti}(2015)}]{Solinas15}%
  \BibitemOpen
  \bibfield  {author} {\bibinfo {author} {\bibfnamefont {P.}~\bibnamefont
  {Solinas}}\ and\ \bibinfo {author} {\bibfnamefont {S.}~\bibnamefont
  {Gasparinetti}},\ }\href@noop {} {\bibfield  {journal} {\bibinfo  {journal}
  {Phys. Rev. E}\ }\textbf {\bibinfo {volume} {92}},\ \bibinfo {pages} {042150}
  (\bibinfo {year} {2015})}\BibitemShut {NoStop}%
\bibitem [{\citenamefont {Kammerlander}\ and\ \citenamefont
  {Anders}(2016)}]{Kammerlander2016}%
  \BibitemOpen
  \bibfield  {author} {\bibinfo {author} {\bibfnamefont {P.}~\bibnamefont
  {Kammerlander}}\ and\ \bibinfo {author} {\bibfnamefont {J.}~\bibnamefont
  {Anders}},\ }\href {\doibase 10.1038/srep22174} {\bibfield  {journal}
  {\bibinfo  {journal} {Scientific reports}\ }\textbf {\bibinfo {volume} {6}},\
  \bibinfo {pages} {22174} (\bibinfo {year} {2016})}\BibitemShut {NoStop}%
\bibitem [{\citenamefont {Deffner}\ \emph {et~al.}(2016)\citenamefont
  {Deffner}, \citenamefont {Paz},\ and\ \citenamefont
  {Zurek}}]{Deffner2016Quantum}%
  \BibitemOpen
  \bibfield  {author} {\bibinfo {author} {\bibfnamefont {S.}~\bibnamefont
  {Deffner}}, \bibinfo {author} {\bibfnamefont {J.~P.}\ \bibnamefont {Paz}}, \
  and\ \bibinfo {author} {\bibfnamefont {W.~H.}\ \bibnamefont {Zurek}},\ }\href
  {\doibase 10.1103/PhysRevE.94.010103} {\bibfield  {journal} {\bibinfo
  {journal} {Phys. Rev. E}\ }\textbf {\bibinfo {volume} {94}},\ \bibinfo
  {pages} {010103} (\bibinfo {year} {2016})}\BibitemShut {NoStop}%
\bibitem [{\citenamefont {Perarnau-Llobet}\ \emph {et~al.}(2017)\citenamefont
  {Perarnau-Llobet}, \citenamefont {B\"aumer}, \citenamefont {Hovhannisyan},
  \citenamefont {Huber},\ and\ \citenamefont {Acin}}]{NoGoTheorem}%
  \BibitemOpen
  \bibfield  {author} {\bibinfo {author} {\bibfnamefont {M.}~\bibnamefont
  {Perarnau-Llobet}}, \bibinfo {author} {\bibfnamefont {E.}~\bibnamefont
  {B\"aumer}}, \bibinfo {author} {\bibfnamefont {K.~V.}\ \bibnamefont
  {Hovhannisyan}}, \bibinfo {author} {\bibfnamefont {M.}~\bibnamefont {Huber}},
  \ and\ \bibinfo {author} {\bibfnamefont {A.}~\bibnamefont {Acin}},\ }\href
  {\doibase 10.1103/PhysRevLett.118.070601} {\bibfield  {journal} {\bibinfo
  {journal} {Phys. Rev. Lett.}\ }\textbf {\bibinfo {volume} {118}},\ \bibinfo
  {pages} {070601} (\bibinfo {year} {2017})}\BibitemShut {NoStop}%
\bibitem [{\citenamefont {Watanabe}\ \emph {et~al.}(2014)\citenamefont
  {Watanabe}, \citenamefont {Venkatesh},\ and\ \citenamefont
  {Talkner}}]{GEMPhysRevE}%
  \BibitemOpen
  \bibfield  {author} {\bibinfo {author} {\bibfnamefont {G.}~\bibnamefont
  {Watanabe}}, \bibinfo {author} {\bibfnamefont {B.~P.}\ \bibnamefont
  {Venkatesh}}, \ and\ \bibinfo {author} {\bibfnamefont {P.}~\bibnamefont
  {Talkner}},\ }\href {\doibase 10.1103/PhysRevE.89.052116} {\bibfield
  {journal} {\bibinfo  {journal} {Phys. Rev. E}\ }\textbf {\bibinfo {volume}
  {89}},\ \bibinfo {pages} {052116} (\bibinfo {year} {2014})}\BibitemShut
  {NoStop}%
\bibitem [{\citenamefont {Venkatesh}\ \emph {et~al.}(2015)\citenamefont
  {Venkatesh}, \citenamefont {Watanabe},\ and\ \citenamefont
  {Talkner}}]{QFTaPMNJP2015}%
  \BibitemOpen
  \bibfield  {author} {\bibinfo {author} {\bibfnamefont {B.~P.}\ \bibnamefont
  {Venkatesh}}, \bibinfo {author} {\bibfnamefont {G.}~\bibnamefont {Watanabe}},
  \ and\ \bibinfo {author} {\bibfnamefont {P.}~\bibnamefont {Talkner}},\ }\href
  {http://stacks.iop.org/1367-2630/17/i=7/a=075018} {\bibfield  {journal}
  {\bibinfo  {journal} {New Journal of Physics}\ }\textbf {\bibinfo {volume}
  {17}},\ \bibinfo {pages} {075018} (\bibinfo {year} {2015})}\BibitemShut
  {NoStop}%
\bibitem [{\citenamefont {Esposito}\ and\ \citenamefont {den
  Broeck}(2011)}]{EspositoSecondLaw}%
  \BibitemOpen
  \bibfield  {author} {\bibinfo {author} {\bibfnamefont {M.}~\bibnamefont
  {Esposito}}\ and\ \bibinfo {author} {\bibfnamefont {C.~V.}\ \bibnamefont {den
  Broeck}},\ }\href {http://stacks.iop.org/0295-5075/95/i=4/a=40004} {\bibfield
   {journal} {\bibinfo  {journal} {EPL (Europhysics Letters)}\ }\textbf
  {\bibinfo {volume} {95}},\ \bibinfo {pages} {40004} (\bibinfo {year}
  {2011})}\BibitemShut {NoStop}%
\bibitem [{\citenamefont {Janzing}(2006)}]{janzing2006quantum}%
  \BibitemOpen
  \bibfield  {author} {\bibinfo {author} {\bibfnamefont {D.}~\bibnamefont
  {Janzing}},\ }\href@noop {} {\bibfield  {journal} {\bibinfo  {journal}
  {Journal of statistical physics}\ }\textbf {\bibinfo {volume} {125}},\
  \bibinfo {pages} {761} (\bibinfo {year} {2006})}\BibitemShut {NoStop}%
\bibitem [{\citenamefont {Lostaglio}\ \emph {et~al.}(2015)\citenamefont
  {Lostaglio}, \citenamefont {Jennings},\ and\ \citenamefont
  {Rudolph}}]{lostaglio2015description}%
  \BibitemOpen
  \bibfield  {author} {\bibinfo {author} {\bibfnamefont {M.}~\bibnamefont
  {Lostaglio}}, \bibinfo {author} {\bibfnamefont {D.}~\bibnamefont {Jennings}},
  \ and\ \bibinfo {author} {\bibfnamefont {T.}~\bibnamefont {Rudolph}},\ }\href
  {\doibase 10.1038/ncomms7383} {\bibfield  {journal} {\bibinfo  {journal}
  {Nat. Commun.}\ }\textbf {\bibinfo {volume} {6}},\ \bibinfo {pages} {6383}
  (\bibinfo {year} {2015})}\BibitemShut {NoStop}%
\bibitem [{\citenamefont {Alonso}\ \emph {et~al.}(2016)\citenamefont {Alonso},
  \citenamefont {Lutz},\ and\ \citenamefont
  {Romito}}]{Alonso2016thermodynamics}%
  \BibitemOpen
  \bibfield  {author} {\bibinfo {author} {\bibfnamefont {J.~J.}\ \bibnamefont
  {Alonso}}, \bibinfo {author} {\bibfnamefont {E.}~\bibnamefont {Lutz}}, \ and\
  \bibinfo {author} {\bibfnamefont {A.}~\bibnamefont {Romito}},\ }\href
  {\doibase 10.1103/PhysRevLett.116.080403} {\bibfield  {journal} {\bibinfo
  {journal} {Phys. Rev. Lett.}\ }\textbf {\bibinfo {volume} {116}},\ \bibinfo
  {pages} {080403} (\bibinfo {year} {2016})}\BibitemShut {NoStop}%
\bibitem [{\citenamefont {Elouard}\ \emph {et~al.}(2017)\citenamefont
  {Elouard}, \citenamefont {Herrera-Mart\'i}, \citenamefont {Clusel},\ and\
  \citenamefont {Auffeves}}]{Elouard2017role}%
  \BibitemOpen
  \bibfield  {author} {\bibinfo {author} {\bibfnamefont {C.}~\bibnamefont
  {Elouard}}, \bibinfo {author} {\bibfnamefont {D.~A.}\ \bibnamefont
  {Herrera-Mart\'i}}, \bibinfo {author} {\bibfnamefont {M.}~\bibnamefont
  {Clusel}}, \ and\ \bibinfo {author} {\bibfnamefont {A.}~\bibnamefont
  {Auffeves}},\ }\href {\doibase 10.1038/s41534-017-0008-4} {\bibfield
  {journal} {\bibinfo  {journal} {npj Quantum Information}\ }\textbf {\bibinfo
  {volume} {3}},\ \bibinfo {pages} {9} (\bibinfo {year} {2017})}\BibitemShut
  {NoStop}%
\bibitem [{\citenamefont {Roncaglia}\ \emph {et~al.}(2014)\citenamefont
  {Roncaglia}, \citenamefont {Cerisola},\ and\ \citenamefont
  {Paz}}]{roncaglia2014work}%
  \BibitemOpen
  \bibfield  {author} {\bibinfo {author} {\bibfnamefont {A.~J.}\ \bibnamefont
  {Roncaglia}}, \bibinfo {author} {\bibfnamefont {F.}~\bibnamefont {Cerisola}},
  \ and\ \bibinfo {author} {\bibfnamefont {J.~P.}\ \bibnamefont {Paz}},\ }\href
  {https://journals.aps.org/prl/abstract/10.1103/PhysRevLett.113.250601}
  {\bibfield  {journal} {\bibinfo  {journal} {Physical review letters}\
  }\textbf {\bibinfo {volume} {113}},\ \bibinfo {pages} {250601} (\bibinfo
  {year} {2014})}\BibitemShut {NoStop}%
\bibitem [{\citenamefont {Chiara}\ \emph {et~al.}(2015)\citenamefont {Chiara},
  \citenamefont {Roncaglia},\ and\ \citenamefont
  {Paz}}]{dechiara2015measuring}%
  \BibitemOpen
  \bibfield  {author} {\bibinfo {author} {\bibfnamefont {G.~D.}\ \bibnamefont
  {Chiara}}, \bibinfo {author} {\bibfnamefont {A.~J.}\ \bibnamefont
  {Roncaglia}}, \ and\ \bibinfo {author} {\bibfnamefont {J.~P.}\ \bibnamefont
  {Paz}},\ }\href {http://stacks.iop.org/1367-2630/17/i=3/a=035004} {\bibfield
  {journal} {\bibinfo  {journal} {New Journal of Physics}\ }\textbf {\bibinfo
  {volume} {17}},\ \bibinfo {pages} {035004} (\bibinfo {year}
  {2015})}\BibitemShut {NoStop}%
\bibitem [{\citenamefont {Talkner}\ and\ \citenamefont
  {H\"anggi}(2016)}]{Talkner2016aspects}%
  \BibitemOpen
  \bibfield  {author} {\bibinfo {author} {\bibfnamefont {P.}~\bibnamefont
  {Talkner}}\ and\ \bibinfo {author} {\bibfnamefont {P.}~\bibnamefont
  {H\"anggi}},\ }\href {\doibase 10.1103/PhysRevE.93.022131} {\bibfield
  {journal} {\bibinfo  {journal} {Phys. Rev. E}\ }\textbf {\bibinfo {volume}
  {93}},\ \bibinfo {pages} {022131} (\bibinfo {year} {2016})}\BibitemShut
  {NoStop}%
\bibitem [{\citenamefont {Solinas}\ \emph {et~al.}(2017)\citenamefont
  {Solinas}, \citenamefont {Miller},\ and\ \citenamefont
  {Anders}}]{Solinas2017measurement}%
  \BibitemOpen
  \bibfield  {author} {\bibinfo {author} {\bibfnamefont {P.}~\bibnamefont
  {Solinas}}, \bibinfo {author} {\bibfnamefont {H.~J.~D.}\ \bibnamefont
  {Miller}}, \ and\ \bibinfo {author} {\bibfnamefont {J.}~\bibnamefont
  {Anders}},\ }\href {\doibase 10.1103/PhysRevA.96.052115} {\bibfield
  {journal} {\bibinfo  {journal} {Phys. Rev. A}\ }\textbf {\bibinfo {volume}
  {96}},\ \bibinfo {pages} {052115} (\bibinfo {year} {2017})}\BibitemShut
  {NoStop}%
\bibitem [{\citenamefont {Hofer}(2017)}]{Hofer2017quasiprobability}%
  \BibitemOpen
  \bibfield  {author} {\bibinfo {author} {\bibfnamefont {P.~P.}\ \bibnamefont
  {Hofer}},\ }\href {\doibase 10.22331/q-2017-10-12-32} {\bibfield  {journal}
  {\bibinfo  {journal} {{Quantum}}\ }\textbf {\bibinfo {volume} {1}},\ \bibinfo
  {pages} {32} (\bibinfo {year} {2017})}\BibitemShut {NoStop}%
\bibitem [{\citenamefont {Sampaio}\ \emph {et~al.}(2018)\citenamefont
  {Sampaio}, \citenamefont {Suomela}, \citenamefont {Ala-Nissila},
  \citenamefont {Anders},\ and\ \citenamefont {Philbin}}]{sampaio17}%
  \BibitemOpen
  \bibfield  {author} {\bibinfo {author} {\bibfnamefont {R.}~\bibnamefont
  {Sampaio}}, \bibinfo {author} {\bibfnamefont {S.}~\bibnamefont {Suomela}},
  \bibinfo {author} {\bibfnamefont {T.}~\bibnamefont {Ala-Nissila}}, \bibinfo
  {author} {\bibfnamefont {J.}~\bibnamefont {Anders}}, \ and\ \bibinfo {author}
  {\bibfnamefont {T.~G.}\ \bibnamefont {Philbin}},\ }\href {\doibase
  10.1103/PhysRevA.97.012131} {\bibfield  {journal} {\bibinfo  {journal} {Phys.
  Rev. A}\ }\textbf {\bibinfo {volume} {97}},\ \bibinfo {pages} {012131}
  (\bibinfo {year} {2018})}\BibitemShut {NoStop}%
\bibitem [{\citenamefont {Lostaglio}(2018)}]{Lostaglio2018}%
  \BibitemOpen
  \bibfield  {author} {\bibinfo {author} {\bibfnamefont {M.}~\bibnamefont
  {Lostaglio}},\ }\href {\doibase 10.1103/PhysRevLett.120.040602} {\bibfield
  {journal} {\bibinfo  {journal} {Phys. Rev. Lett.}\ }\textbf {\bibinfo
  {volume} {120}},\ \bibinfo {pages} {040602} (\bibinfo {year}
  {2018})}\BibitemShut {NoStop}%
\bibitem [{\citenamefont {Alhambra}\ \emph {et~al.}(2016)\citenamefont
  {Alhambra}, \citenamefont {Masanes}, \citenamefont {Oppenheim},\ and\
  \citenamefont {Perry}}]{Alhambra2016}%
  \BibitemOpen
  \bibfield  {author} {\bibinfo {author} {\bibfnamefont {A.~M.}\ \bibnamefont
  {Alhambra}}, \bibinfo {author} {\bibfnamefont {L.}~\bibnamefont {Masanes}},
  \bibinfo {author} {\bibfnamefont {J.}~\bibnamefont {Oppenheim}}, \ and\
  \bibinfo {author} {\bibfnamefont {C.}~\bibnamefont {Perry}},\ }\href
  {\doibase 10.1103/PhysRevX.6.041017} {\bibfield  {journal} {\bibinfo
  {journal} {Phys. Rev. X}\ }\textbf {\bibinfo {volume} {6}},\ \bibinfo {pages}
  {041017} (\bibinfo {year} {2016})}\BibitemShut {NoStop}%
\bibitem [{\citenamefont {Richens}\ and\ \citenamefont
  {Masanes}(2016)}]{richens2016work}%
  \BibitemOpen
  \bibfield  {author} {\bibinfo {author} {\bibfnamefont {J.~G.}\ \bibnamefont
  {Richens}}\ and\ \bibinfo {author} {\bibfnamefont {L.}~\bibnamefont
  {Masanes}},\ }\href@noop {} {\bibfield  {journal} {\bibinfo  {journal}
  {Nature communications}\ }\textbf {\bibinfo {volume} {7}},\ \bibinfo {pages}
  {13511} (\bibinfo {year} {2016})}\BibitemShut {NoStop}%
\bibitem [{\citenamefont {\AA{}berg}(2018)}]{Aberg2018}%
  \BibitemOpen
  \bibfield  {author} {\bibinfo {author} {\bibfnamefont {J.}~\bibnamefont
  {\AA{}berg}},\ }\href {\doibase 10.1103/PhysRevX.8.011019} {\bibfield
  {journal} {\bibinfo  {journal} {Phys. Rev. X}\ }\textbf {\bibinfo {volume}
  {8}},\ \bibinfo {pages} {011019} (\bibinfo {year} {2018})}\BibitemShut
  {NoStop}%
\bibitem [{\citenamefont {Ng}\ and\ \citenamefont
  {Woods}(2018)}]{nelly2018quantum}%
  \BibitemOpen
  \bibfield  {author} {\bibinfo {author} {\bibfnamefont {N.}~\bibnamefont
  {Ng}}\ and\ \bibinfo {author} {\bibfnamefont {M.~P.}\ \bibnamefont {Woods}},\
  }\href {https://arxiv.org/abs/1805.09564} {\bibfield  {journal} {\bibinfo
  {journal} {arXiv:1805.09564}\ } (\bibinfo {year} {2018})}\BibitemShut
  {NoStop}%
\bibitem [{\citenamefont {Bell}(1966)}]{bell1966problem}%
  \BibitemOpen
  \bibfield  {author} {\bibinfo {author} {\bibfnamefont {J.~S.}\ \bibnamefont
  {Bell}},\ }\href@noop {} {\bibfield  {journal} {\bibinfo  {journal} {Rev.
  Mod. Phys.}\ }\textbf {\bibinfo {volume} {38}},\ \bibinfo {pages} {447}
  (\bibinfo {year} {1966})}\BibitemShut {NoStop}%
\bibitem [{\citenamefont {Kochen}\ and\ \citenamefont
  {Specker}(1975)}]{kochen1975problem}%
  \BibitemOpen
  \bibfield  {author} {\bibinfo {author} {\bibfnamefont {S.}~\bibnamefont
  {Kochen}}\ and\ \bibinfo {author} {\bibfnamefont {E.~P.}\ \bibnamefont
  {Specker}},\ }in\ \href@noop {} {\emph {\bibinfo {booktitle} {The
  Logico-Algebraic Approach to Quantum Mechanics}}}\ (\bibinfo  {publisher}
  {Springer},\ \bibinfo {year} {1975})\ pp.\ \bibinfo {pages}
  {293--328}\BibitemShut {NoStop}%
\bibitem [{\citenamefont {Howard}\ \emph {et~al.}(2014)\citenamefont {Howard},
  \citenamefont {Wallman}, \citenamefont {Veitch},\ and\ \citenamefont
  {Emerson}}]{howard2014contextuality}%
  \BibitemOpen
  \bibfield  {author} {\bibinfo {author} {\bibfnamefont {M.}~\bibnamefont
  {Howard}}, \bibinfo {author} {\bibfnamefont {J.}~\bibnamefont {Wallman}},
  \bibinfo {author} {\bibfnamefont {V.}~\bibnamefont {Veitch}}, \ and\ \bibinfo
  {author} {\bibfnamefont {J.}~\bibnamefont {Emerson}},\ }\href
  {https://www.nature.com/articles/nature13460} {\bibfield  {journal} {\bibinfo
   {journal} {Nature}\ }\textbf {\bibinfo {volume} {510}},\ \bibinfo {pages}
  {351} (\bibinfo {year} {2014})}\BibitemShut {NoStop}%
\bibitem [{\citenamefont {Delfosse}\ \emph {et~al.}(2015)\citenamefont
  {Delfosse}, \citenamefont {Guerin}, \citenamefont {Bian},\ and\ \citenamefont
  {Raussendorf}}]{delfosse2015wigner}%
  \BibitemOpen
  \bibfield  {author} {\bibinfo {author} {\bibfnamefont {N.}~\bibnamefont
  {Delfosse}}, \bibinfo {author} {\bibfnamefont {P.~A.}\ \bibnamefont
  {Guerin}}, \bibinfo {author} {\bibfnamefont {J.}~\bibnamefont {Bian}}, \ and\
  \bibinfo {author} {\bibfnamefont {R.}~\bibnamefont {Raussendorf}},\ }\href
  {https://journals.aps.org/prx/abstract/10.1103/PhysRevX.5.021003} {\bibfield
  {journal} {\bibinfo  {journal} {Phys. Rev.~X}\ }\textbf {\bibinfo {volume}
  {5}},\ \bibinfo {pages} {021003} (\bibinfo {year} {2015})}\BibitemShut
  {NoStop}%
\bibitem [{\citenamefont {Bermejo-Vega}\ \emph {et~al.}(2017)\citenamefont
  {Bermejo-Vega}, \citenamefont {Delfosse}, \citenamefont {Browne},
  \citenamefont {Okay},\ and\ \citenamefont
  {Raussendorf}}]{bermejo2016contextuality}%
  \BibitemOpen
  \bibfield  {author} {\bibinfo {author} {\bibfnamefont {J.}~\bibnamefont
  {Bermejo-Vega}}, \bibinfo {author} {\bibfnamefont {N.}~\bibnamefont
  {Delfosse}}, \bibinfo {author} {\bibfnamefont {D.~E.}\ \bibnamefont
  {Browne}}, \bibinfo {author} {\bibfnamefont {C.}~\bibnamefont {Okay}}, \ and\
  \bibinfo {author} {\bibfnamefont {R.}~\bibnamefont {Raussendorf}},\ }\href
  {\doibase 10.1103/PhysRevLett.119.120505} {\bibfield  {journal} {\bibinfo
  {journal} {Phys. Rev. Lett.}\ }\textbf {\bibinfo {volume} {119}},\ \bibinfo
  {pages} {120505} (\bibinfo {year} {2017})}\BibitemShut {NoStop}%
\bibitem [{\citenamefont {Spekkens}(2005)}]{spekkens2005contextuality}%
  \BibitemOpen
  \bibfield  {author} {\bibinfo {author} {\bibfnamefont {R.~W.}\ \bibnamefont
  {Spekkens}},\ }\href
  {https://journals.aps.org/pra/abstract/10.1103/PhysRevA.71.052108} {\bibfield
   {journal} {\bibinfo  {journal} {Phys. Rev. A}\ }\textbf {\bibinfo {volume}
  {71}},\ \bibinfo {pages} {052108} (\bibinfo {year} {2005})}\BibitemShut
  {NoStop}%
\bibitem [{\citenamefont {Brunner}\ \emph
  {et~al.}(2014{\natexlab{b}})\citenamefont {Brunner}, \citenamefont
  {Cavalcanti}, \citenamefont {Pironio}, \citenamefont {Scarani},\ and\
  \citenamefont {Wehner}}]{brunner2014bell}%
  \BibitemOpen
  \bibfield  {author} {\bibinfo {author} {\bibfnamefont {N.}~\bibnamefont
  {Brunner}}, \bibinfo {author} {\bibfnamefont {D.}~\bibnamefont {Cavalcanti}},
  \bibinfo {author} {\bibfnamefont {S.}~\bibnamefont {Pironio}}, \bibinfo
  {author} {\bibfnamefont {V.}~\bibnamefont {Scarani}}, \ and\ \bibinfo
  {author} {\bibfnamefont {S.}~\bibnamefont {Wehner}},\ }\href
  {https://journals.aps.org/rmp/abstract/10.1103/RevModPhys.86.419} {\bibfield
  {journal} {\bibinfo  {journal} {Rev. Mod. Phys.}\ }\textbf {\bibinfo {volume}
  {86}},\ \bibinfo {pages} {419} (\bibinfo {year}
  {2014}{\natexlab{b}})}\BibitemShut {NoStop}%
\bibitem [{\citenamefont {Solinas}\ and\ \citenamefont
  {Gasparinetti}(2016)}]{Solinas2016probing}%
  \BibitemOpen
  \bibfield  {author} {\bibinfo {author} {\bibfnamefont {P.}~\bibnamefont
  {Solinas}}\ and\ \bibinfo {author} {\bibfnamefont {S.}~\bibnamefont
  {Gasparinetti}},\ }\href {\doibase 10.1103/PhysRevA.94.052103} {\bibfield
  {journal} {\bibinfo  {journal} {Phys. Rev. A}\ }\textbf {\bibinfo {volume}
  {94}},\ \bibinfo {pages} {052103} (\bibinfo {year} {2016})}\BibitemShut
  {NoStop}%
\bibitem [{\citenamefont {Xu}\ \emph {et~al.}(2018)\citenamefont {Xu},
  \citenamefont {Zou}, \citenamefont {Guo},\ and\ \citenamefont
  {Kong}}]{xu2017effects}%
  \BibitemOpen
  \bibfield  {author} {\bibinfo {author} {\bibfnamefont {B.-M.}\ \bibnamefont
  {Xu}}, \bibinfo {author} {\bibfnamefont {J.}~\bibnamefont {Zou}}, \bibinfo
  {author} {\bibfnamefont {L.-S.}\ \bibnamefont {Guo}}, \ and\ \bibinfo
  {author} {\bibfnamefont {X.-M.}\ \bibnamefont {Kong}},\ }\href {\doibase
  10.1103/PhysRevA.97.052122} {\bibfield  {journal} {\bibinfo  {journal} {Phys.
  Rev. A}\ }\textbf {\bibinfo {volume} {97}},\ \bibinfo {pages} {052122}
  (\bibinfo {year} {2018})}\BibitemShut {NoStop}%
\bibitem [{\citenamefont {Aharonov}\ \emph
  {et~al.}(1988{\natexlab{a}})\citenamefont {Aharonov}, \citenamefont
  {Albert},\ and\ \citenamefont {Vaidman}}]{aharonov1988result}%
  \BibitemOpen
  \bibfield  {author} {\bibinfo {author} {\bibfnamefont {Y.}~\bibnamefont
  {Aharonov}}, \bibinfo {author} {\bibfnamefont {D.~Z.}\ \bibnamefont
  {Albert}}, \ and\ \bibinfo {author} {\bibfnamefont {L.}~\bibnamefont
  {Vaidman}},\ }\href {\doibase 10.1103/PhysRevLett.60.1351} {\bibfield
  {journal} {\bibinfo  {journal} {Phys. Rev. Lett.}\ }\textbf {\bibinfo
  {volume} {60}},\ \bibinfo {pages} {1351} (\bibinfo {year}
  {1988}{\natexlab{a}})}\BibitemShut {NoStop}%
\bibitem [{\citenamefont {Wiseman}(2002)}]{wiseman2002weak}%
  \BibitemOpen
  \bibfield  {author} {\bibinfo {author} {\bibfnamefont {H.~M.}\ \bibnamefont
  {Wiseman}},\ }\href {\doibase 10.1103/PhysRevA.65.032111} {\bibfield
  {journal} {\bibinfo  {journal} {Phys. Rev. A}\ }\textbf {\bibinfo {volume}
  {65}},\ \bibinfo {pages} {032111} (\bibinfo {year} {2002})}\BibitemShut
  {NoStop}%
\bibitem [{\citenamefont {Miller}\ and\ \citenamefont
  {Anders}(2017)}]{Miller2017time}%
  \BibitemOpen
  \bibfield  {author} {\bibinfo {author} {\bibfnamefont {H.~J.~D.}\
  \bibnamefont {Miller}}\ and\ \bibinfo {author} {\bibfnamefont
  {J.}~\bibnamefont {Anders}},\ }\href
  {http://stacks.iop.org/1367-2630/19/i=6/a=062001} {\bibfield  {journal}
  {\bibinfo  {journal} {New Journal of Physics}\ }\textbf {\bibinfo {volume}
  {19}},\ \bibinfo {pages} {062001} (\bibinfo {year} {2017})}\BibitemShut
  {NoStop}%
\bibitem [{\citenamefont {Hall}(2004)}]{hall2004prior}%
  \BibitemOpen
  \bibfield  {author} {\bibinfo {author} {\bibfnamefont {J.~W.}\ \bibnamefont
  {Hall}},\ }\href {\doibase 10.1103/PhysRevA.69.052113} {\bibfield  {journal}
  {\bibinfo  {journal} {Phys. Rev. A}\ }\textbf {\bibinfo {volume} {69}},\
  \bibinfo {pages} {052113} (\bibinfo {year} {2004})}\BibitemShut {NoStop}%
\bibitem [{\citenamefont {Griffiths}(1984)}]{griffiths1984consistent}%
  \BibitemOpen
  \bibfield  {author} {\bibinfo {author} {\bibfnamefont {R.~B.}\ \bibnamefont
  {Griffiths}},\ }\href@noop {} {\bibfield  {journal} {\bibinfo  {journal}
  {Journal of Statistical Physics}\ }\textbf {\bibinfo {volume} {36}},\
  \bibinfo {pages} {219} (\bibinfo {year} {1984})}\BibitemShut {NoStop}%
\bibitem [{\citenamefont {Goldstein}\ and\ \citenamefont
  {Page}(1995)}]{goldstein1995linearly}%
  \BibitemOpen
  \bibfield  {author} {\bibinfo {author} {\bibfnamefont {S.}~\bibnamefont
  {Goldstein}}\ and\ \bibinfo {author} {\bibfnamefont {D.~N.}\ \bibnamefont
  {Page}},\ }\href
  {https://journals.aps.org/prl/abstract/10.1103/PhysRevLett.74.3715}
  {\bibfield  {journal} {\bibinfo  {journal} {Physical review letters}\
  }\textbf {\bibinfo {volume} {74}},\ \bibinfo {pages} {3715} (\bibinfo {year}
  {1995})}\BibitemShut {NoStop}%
\bibitem [{\citenamefont {Sagawa}(2012)}]{SAGAWA2012}%
  \BibitemOpen
  \bibfield  {author} {\bibinfo {author} {\bibfnamefont {T.}~\bibnamefont
  {Sagawa}},\ }\enquote {\bibinfo {title} {Second law-like inequalities with
  quantum relative entropy: An introduction},}\ in\ \href {\doibase
  10.1142/9789814425193_0003} {\emph {\bibinfo {booktitle} {Lectures on Quantum
  Computing, Thermodynamics and Statistical Physics}}}\ (\bibinfo  {publisher}
  {World Scientific},\ \bibinfo {year} {2012})\ pp.\ \bibinfo {pages}
  {125--190}\BibitemShut {NoStop}%
\bibitem [{\citenamefont
  {Allahverdyan}(2014{\natexlab{b}})}]{Allahverdyan2014nonequilibrium}%
  \BibitemOpen
  \bibfield  {author} {\bibinfo {author} {\bibfnamefont {A.~E.}\ \bibnamefont
  {Allahverdyan}},\ }\href {\doibase 10.1103/PhysRevE.90.032137} {\bibfield
  {journal} {\bibinfo  {journal} {Phys. Rev. E}\ }\textbf {\bibinfo {volume}
  {90}},\ \bibinfo {pages} {032137} (\bibinfo {year}
  {2014}{\natexlab{b}})}\BibitemShut {NoStop}%
\bibitem [{\citenamefont {De~Chiara}\ \emph {et~al.}(2018)\citenamefont
  {De~Chiara}, \citenamefont {Solinas}, \citenamefont {Cerisola},\ and\
  \citenamefont {Roncaglia}}]{de2018ancilla}%
  \BibitemOpen
  \bibfield  {author} {\bibinfo {author} {\bibfnamefont {G.}~\bibnamefont
  {De~Chiara}}, \bibinfo {author} {\bibfnamefont {P.}~\bibnamefont {Solinas}},
  \bibinfo {author} {\bibfnamefont {F.}~\bibnamefont {Cerisola}}, \ and\
  \bibinfo {author} {\bibfnamefont {A.~J.}\ \bibnamefont {Roncaglia}},\ }\href
  {https://arxiv.org/abs/1805.06047} {\bibfield  {journal} {\bibinfo  {journal}
  {arXiv:1805.06047}\ } (\bibinfo {year} {2018})}\BibitemShut {NoStop}%
\bibitem [{\citenamefont {Pusey}(2014)}]{pusey2014anomalous}%
  \BibitemOpen
  \bibfield  {author} {\bibinfo {author} {\bibfnamefont {M.~F.}\ \bibnamefont
  {Pusey}},\ }\href {\doibase 10.1103/PhysRevLett.113.200401} {\bibfield
  {journal} {\bibinfo  {journal} {Phys. Rev. Lett.}\ }\textbf {\bibinfo
  {volume} {113}},\ \bibinfo {pages} {200401} (\bibinfo {year}
  {2014})}\BibitemShut {NoStop}%
\bibitem [{\citenamefont {Williams}\ and\ \citenamefont
  {Jordan}(2008)}]{Williams2008}%
  \BibitemOpen
  \bibfield  {author} {\bibinfo {author} {\bibfnamefont {N.~S.}\ \bibnamefont
  {Williams}}\ and\ \bibinfo {author} {\bibfnamefont {A.~N.}\ \bibnamefont
  {Jordan}},\ }\href {\doibase 10.1103/PhysRevLett.100.026804} {\bibfield
  {journal} {\bibinfo  {journal} {Phys. Rev. Lett.}\ }\textbf {\bibinfo
  {volume} {100}},\ \bibinfo {pages} {026804} (\bibinfo {year}
  {2008})}\BibitemShut {NoStop}%
\bibitem [{\citenamefont {Blattmann}\ and\ \citenamefont
  {M{\o}lmer}(2017)}]{blattmann2017macroscopic}%
  \BibitemOpen
  \bibfield  {author} {\bibinfo {author} {\bibfnamefont {R.}~\bibnamefont
  {Blattmann}}\ and\ \bibinfo {author} {\bibfnamefont {K.}~\bibnamefont
  {M{\o}lmer}},\ }\href
  {https://journals.aps.org/pra/abstract/10.1103/PhysRevA.96.012115} {\bibfield
   {journal} {\bibinfo  {journal} {Physical Review A}\ }\textbf {\bibinfo
  {volume} {96}},\ \bibinfo {pages} {012115} (\bibinfo {year}
  {2017})}\BibitemShut {NoStop}%
\bibitem [{\citenamefont {Miller}\ and\ \citenamefont
  {Anders}(2018)}]{Miller2017Legget}%
  \BibitemOpen
  \bibfield  {author} {\bibinfo {author} {\bibfnamefont {H.~J.~D.}\
  \bibnamefont {Miller}}\ and\ \bibinfo {author} {\bibfnamefont
  {J.}~\bibnamefont {Anders}},\ }\href@noop {} {\bibfield  {journal} {\bibinfo
  {journal} {Entropy}\ }\textbf {\bibinfo {volume} {20}} (\bibinfo {year}
  {2018})}\BibitemShut {NoStop}%
\bibitem [{\citenamefont {Bohm}(1952{\natexlab{a}})}]{Bohm:1952aa}%
  \BibitemOpen
  \bibfield  {author} {\bibinfo {author} {\bibfnamefont {D.}~\bibnamefont
  {Bohm}},\ }\href {\doibase 10.1103/PhysRev.85.166} {\bibfield  {journal}
  {\bibinfo  {journal} {Physical Review}\ }\textbf {\bibinfo {volume} {85}},\
  \bibinfo {pages} {166} (\bibinfo {year} {1952}{\natexlab{a}})}\BibitemShut
  {NoStop}%
\bibitem [{\citenamefont {Bohm}(1952{\natexlab{b}})}]{Bohm:1952ab}%
  \BibitemOpen
  \bibfield  {author} {\bibinfo {author} {\bibfnamefont {D.}~\bibnamefont
  {Bohm}},\ }\href {\doibase 10.1103/PhysRev.85.180} {\bibfield  {journal}
  {\bibinfo  {journal} {Physical Review}\ }\textbf {\bibinfo {volume} {85}},\
  \bibinfo {pages} {180} (\bibinfo {year} {1952}{\natexlab{b}})}\BibitemShut
  {NoStop}%
\bibitem [{\citenamefont {Broglie}(1956)}]{broglie1956tentative}%
  \BibitemOpen
  \bibfield  {author} {\bibinfo {author} {\bibfnamefont {L.}~\bibnamefont
  {Broglie}},\ }\href@noop {} {\emph {\bibinfo {title} {Une tentative
  d'interpretation causale et non lineaire de la mecanique ondulatoire (la
  theorie de la double solution).}}}\ (\bibinfo  {publisher}
  {Gauthier-Villars},\ \bibinfo {year} {1956})\BibitemShut {NoStop}%
\bibitem [{\citenamefont {Holland}(1995)}]{holland1995quantum}%
  \BibitemOpen
  \bibfield  {author} {\bibinfo {author} {\bibfnamefont {P.~R.}\ \bibnamefont
  {Holland}},\ }\href@noop {} {\emph {\bibinfo {title} {The quantum theory of
  motion: an account of the de Broglie-Bohm causal interpretation of quantum
  mechanics}}}\ (\bibinfo  {publisher} {Cambridge university press},\ \bibinfo
  {year} {1995})\BibitemShut {NoStop}%
\bibitem [{\citenamefont {D{\"u}rr}\ \emph {et~al.}(2013)\citenamefont
  {D{\"u}rr}, \citenamefont {Goldstein},\ and\ \citenamefont
  {Zangh{\`\i}}}]{durr2012quantum}%
  \BibitemOpen
  \bibfield  {author} {\bibinfo {author} {\bibfnamefont {D.}~\bibnamefont
  {D{\"u}rr}}, \bibinfo {author} {\bibfnamefont {S.}~\bibnamefont {Goldstein}},
  \ and\ \bibinfo {author} {\bibfnamefont {N.}~\bibnamefont {Zangh{\`\i}}},\
  }\href@noop {} {\emph {\bibinfo {title} {{Quantum physics without quantum
  philosophy}}}}\ (\bibinfo  {publisher} {Springer Science \& Business Media},\
  \bibinfo {year} {2013})\BibitemShut {NoStop}%
\bibitem [{\citenamefont {Aharonov}\ \emph
  {et~al.}(1988{\natexlab{b}})\citenamefont {Aharonov}, \citenamefont
  {Albert},\ and\ \citenamefont {Vaidman}}]{AAV88}%
  \BibitemOpen
  \bibfield  {author} {\bibinfo {author} {\bibfnamefont {Y.}~\bibnamefont
  {Aharonov}}, \bibinfo {author} {\bibfnamefont {D.~Z.}\ \bibnamefont
  {Albert}}, \ and\ \bibinfo {author} {\bibfnamefont {L.}~\bibnamefont
  {Vaidman}},\ }\href {\doibase 10.1103/PhysRevLett.60.1351} {\bibfield
  {journal} {\bibinfo  {journal} {Phys. Rev. Lett.}\ }\textbf {\bibinfo
  {volume} {60}},\ \bibinfo {pages} {1351} (\bibinfo {year}
  {1988}{\natexlab{b}})}\BibitemShut {NoStop}%
\bibitem [{\citenamefont {Duck}\ \emph {et~al.}(1989)\citenamefont {Duck},
  \citenamefont {Stevenson},\ and\ \citenamefont {Sudarshan}}]{Duck89}%
  \BibitemOpen
  \bibfield  {author} {\bibinfo {author} {\bibfnamefont {I.~M.}\ \bibnamefont
  {Duck}}, \bibinfo {author} {\bibfnamefont {P.~M.}\ \bibnamefont {Stevenson}},
  \ and\ \bibinfo {author} {\bibfnamefont {E.~C.~G.}\ \bibnamefont
  {Sudarshan}},\ }\href {\doibase 10.1103/PhysRevD.40.2112} {\bibfield
  {journal} {\bibinfo  {journal} {Phys. Rev. D}\ }\textbf {\bibinfo {volume}
  {40}},\ \bibinfo {pages} {2112} (\bibinfo {year} {1989})}\BibitemShut
  {NoStop}%
\bibitem [{\citenamefont {Kocsis}\ \emph {et~al.}(2011)\citenamefont {Kocsis},
  \citenamefont {Braverman}, \citenamefont {Ravets}, \citenamefont {Stevens},
  \citenamefont {Mirin}, \citenamefont {Shalm},\ and\ \citenamefont
  {Steinberg}}]{Kocsis2011}%
  \BibitemOpen
  \bibfield  {author} {\bibinfo {author} {\bibfnamefont {S.}~\bibnamefont
  {Kocsis}}, \bibinfo {author} {\bibfnamefont {B.}~\bibnamefont {Braverman}},
  \bibinfo {author} {\bibfnamefont {S.}~\bibnamefont {Ravets}}, \bibinfo
  {author} {\bibfnamefont {M.~J.}\ \bibnamefont {Stevens}}, \bibinfo {author}
  {\bibfnamefont {R.~P.}\ \bibnamefont {Mirin}}, \bibinfo {author}
  {\bibfnamefont {L.~K.}\ \bibnamefont {Shalm}}, \ and\ \bibinfo {author}
  {\bibfnamefont {A.~M.}\ \bibnamefont {Steinberg}},\ }\href {\doibase
  10.1126/science.1202218} {\bibfield  {journal} {\bibinfo  {journal}
  {Science}\ }\textbf {\bibinfo {volume} {332}},\ \bibinfo {pages} {1170}
  (\bibinfo {year} {2011})}\BibitemShut {NoStop}%
\bibitem [{\citenamefont {Mahler}\ \emph {et~al.}(2016)\citenamefont {Mahler},
  \citenamefont {Rozema}, \citenamefont {Fisher}, \citenamefont {Vermeyden},
  \citenamefont {Resch}, \citenamefont {Wiseman},\ and\ \citenamefont
  {Steinberg}}]{Mahler2016}%
  \BibitemOpen
  \bibfield  {author} {\bibinfo {author} {\bibfnamefont {D.~H.}\ \bibnamefont
  {Mahler}}, \bibinfo {author} {\bibfnamefont {L.}~\bibnamefont {Rozema}},
  \bibinfo {author} {\bibfnamefont {K.}~\bibnamefont {Fisher}}, \bibinfo
  {author} {\bibfnamefont {L.}~\bibnamefont {Vermeyden}}, \bibinfo {author}
  {\bibfnamefont {K.~J.}\ \bibnamefont {Resch}}, \bibinfo {author}
  {\bibfnamefont {H.~M.}\ \bibnamefont {Wiseman}}, \ and\ \bibinfo {author}
  {\bibfnamefont {A.}~\bibnamefont {Steinberg}},\ }\href {\doibase
  10.1126/sciadv.1501466} {\bibfield  {journal} {\bibinfo  {journal} {Science
  Advances}\ }\textbf {\bibinfo {volume} {2}},\ \bibinfo {pages} {e1501466}
  (\bibinfo {year} {2016})}\BibitemShut {NoStop}%
\bibitem [{\citenamefont {Xiao}\ \emph {et~al.}(2017)\citenamefont {Xiao},
  \citenamefont {Kedem}, \citenamefont {Xu}, \citenamefont {Li},\ and\
  \citenamefont {Guo}}]{Xiao2017}%
  \BibitemOpen
  \bibfield  {author} {\bibinfo {author} {\bibfnamefont {Y.}~\bibnamefont
  {Xiao}}, \bibinfo {author} {\bibfnamefont {Y.}~\bibnamefont {Kedem}},
  \bibinfo {author} {\bibfnamefont {J.-S.}\ \bibnamefont {Xu}}, \bibinfo
  {author} {\bibfnamefont {C.-F.}\ \bibnamefont {Li}}, \ and\ \bibinfo {author}
  {\bibfnamefont {G.-C.}\ \bibnamefont {Guo}},\ }\href {\doibase
  10.1364/OE.25.014463} {\bibfield  {journal} {\bibinfo  {journal} {Opt.
  Express}\ }\textbf {\bibinfo {volume} {25}},\ \bibinfo {pages} {14463}
  (\bibinfo {year} {2017})}\BibitemShut {NoStop}%
\bibitem [{\citenamefont {Wu}\ \emph {et~al.}()\citenamefont {Wu},
  \citenamefont {Yuan}, \citenamefont {Xiang}, \citenamefont {Li},
  \citenamefont {Guo},\ and\ \citenamefont {Perarnau-Llobet}}]{kangda2018}%
  \BibitemOpen
  \bibfield  {author} {\bibinfo {author} {\bibfnamefont {K.-D.}\ \bibnamefont
  {Wu}}, \bibinfo {author} {\bibfnamefont {Y.}~\bibnamefont {Yuan}}, \bibinfo
  {author} {\bibfnamefont {G.-Y.}\ \bibnamefont {Xiang}}, \bibinfo {author}
  {\bibfnamefont {C.-F.}\ \bibnamefont {Li}}, \bibinfo {author} {\bibfnamefont
  {G.-C.}\ \bibnamefont {Guo}}, \ and\ \bibinfo {author} {\bibfnamefont
  {M.}~\bibnamefont {Perarnau-Llobet}},\ }\href@noop {} {\bibinfo  {journal}
  {in prep.}\ }\BibitemShut {NoStop}%
\bibitem [{\citenamefont {Pashayan}\ \emph {et~al.}(2015)\citenamefont
  {Pashayan}, \citenamefont {Wallman},\ and\ \citenamefont
  {Bartlett}}]{pashayan2015estimating}%
  \BibitemOpen
\bibfield  {journal} {  }\bibfield  {author} {\bibinfo {author} {\bibfnamefont
  {H.}~\bibnamefont {Pashayan}}, \bibinfo {author} {\bibfnamefont {J.~J.}\
  \bibnamefont {Wallman}}, \ and\ \bibinfo {author} {\bibfnamefont {S.~D.}\
  \bibnamefont {Bartlett}},\ }\href
  {https://journals.aps.org/prl/abstract/10.1103/PhysRevLett.115.070501}
  {\bibfield  {journal} {\bibinfo  {journal} {Physical review letters}\
  }\textbf {\bibinfo {volume} {115}},\ \bibinfo {pages} {070501} (\bibinfo
  {year} {2015})}\BibitemShut {NoStop}%
\bibitem [{\citenamefont {Levy}\ and\ \citenamefont
  {Gelbwaser-Klimovsky}(2018{\natexlab{b}})}]{Amikam2018}%
  \BibitemOpen
  \bibfield  {author} {\bibinfo {author} {\bibfnamefont {A.}~\bibnamefont
  {Levy}}\ and\ \bibinfo {author} {\bibfnamefont {D.}~\bibnamefont
  {Gelbwaser-Klimovsky}},\ }\href {https://arxiv.org/abs/1803.05586} {\bibfield
   {journal} {\bibinfo  {journal} {arXiv:1803.05586}\ } (\bibinfo {year}
  {2018}{\natexlab{b}})}\BibitemShut {NoStop}%
\bibitem [{\citenamefont {Pironio}\ \emph {et~al.}(2010)\citenamefont
  {Pironio}, \citenamefont {Ac{\'\i}n}, \citenamefont {Massar}, \citenamefont
  {de~La~Giroday}, \citenamefont {Matsukevich}, \citenamefont {Maunz},
  \citenamefont {Olmschenk}, \citenamefont {Hayes}, \citenamefont {Luo},
  \citenamefont {Manning} \emph {et~al.}}]{pironio2010random}%
  \BibitemOpen
  \bibfield  {author} {\bibinfo {author} {\bibfnamefont {S.}~\bibnamefont
  {Pironio}}, \bibinfo {author} {\bibfnamefont {A.}~\bibnamefont {Ac{\'\i}n}},
  \bibinfo {author} {\bibfnamefont {S.}~\bibnamefont {Massar}}, \bibinfo
  {author} {\bibfnamefont {A.~B.}\ \bibnamefont {de~La~Giroday}}, \bibinfo
  {author} {\bibfnamefont {D.~N.}\ \bibnamefont {Matsukevich}}, \bibinfo
  {author} {\bibfnamefont {P.}~\bibnamefont {Maunz}}, \bibinfo {author}
  {\bibfnamefont {S.}~\bibnamefont {Olmschenk}}, \bibinfo {author}
  {\bibfnamefont {D.}~\bibnamefont {Hayes}}, \bibinfo {author} {\bibfnamefont
  {L.}~\bibnamefont {Luo}}, \bibinfo {author} {\bibfnamefont {T.~A.}\
  \bibnamefont {Manning}},  \emph {et~al.},\ }\href
  {https://www.nature.com/articles/nature09008} {\bibfield  {journal} {\bibinfo
   {journal} {Nature}\ }\textbf {\bibinfo {volume} {464}},\ \bibinfo {pages}
  {1021} (\bibinfo {year} {2010})}\BibitemShut {NoStop}%
\bibitem [{\citenamefont {Mazurek}\ \emph {et~al.}(2016)\citenamefont
  {Mazurek}, \citenamefont {Pusey}, \citenamefont {Kunjwal}, \citenamefont
  {Resch},\ and\ \citenamefont {Spekkens}}]{mazurek2016experimental}%
  \BibitemOpen
  \bibfield  {author} {\bibinfo {author} {\bibfnamefont {M.~D.}\ \bibnamefont
  {Mazurek}}, \bibinfo {author} {\bibfnamefont {M.~F.}\ \bibnamefont {Pusey}},
  \bibinfo {author} {\bibfnamefont {R.}~\bibnamefont {Kunjwal}}, \bibinfo
  {author} {\bibfnamefont {K.~J.}\ \bibnamefont {Resch}}, \ and\ \bibinfo
  {author} {\bibfnamefont {R.~W.}\ \bibnamefont {Spekkens}},\ }\href
  {https://www.nature.com/articles/ncomms11780} {\bibfield  {journal} {\bibinfo
   {journal} {Nature communications}\ }\textbf {\bibinfo {volume} {7}},\
  \bibinfo {pages} {ncomms11780} (\bibinfo {year} {2016})}\BibitemShut
  {NoStop}%
\bibitem [{\citenamefont {{Skrzypczyk}}\ \emph {et~al.}(2014)\citenamefont
  {{Skrzypczyk}}, \citenamefont {{Short}},\ and\ \citenamefont
  {{Popescu}}}]{skrzypczyk2014work}%
  \BibitemOpen
  \bibfield  {author} {\bibinfo {author} {\bibfnamefont {P.}~\bibnamefont
  {{Skrzypczyk}}}, \bibinfo {author} {\bibfnamefont {A.~J.}\ \bibnamefont
  {{Short}}}, \ and\ \bibinfo {author} {\bibfnamefont {S.}~\bibnamefont
  {{Popescu}}},\ }\href {\doibase 10.1038/ncomms5185} {\bibfield  {journal}
  {\bibinfo  {journal} {Nature Communications}\ }\textbf {\bibinfo {volume}
  {5}},\ \bibinfo {eid} {4185} (\bibinfo {year} {2014})},\ \Eprint
  {http://arxiv.org/abs/1307.1558} {arXiv:1307.1558 [quant-ph]} \BibitemShut
  {NoStop}%
\bibitem [{\citenamefont {\AA{}berg}(2014)}]{aberg2014catalytic}%
  \BibitemOpen
  \bibfield  {author} {\bibinfo {author} {\bibfnamefont {J.}~\bibnamefont
  {\AA{}berg}},\ }\href {\doibase 10.1103/PhysRevLett.113.150402} {\bibfield
  {journal} {\bibinfo  {journal} {Phys. Rev. Lett.}\ }\textbf {\bibinfo
  {volume} {113}},\ \bibinfo {pages} {150402} (\bibinfo {year}
  {2014})}\BibitemShut {NoStop}%
\bibitem [{\citenamefont {Reeb}\ and\ \citenamefont {Wolf}(2014)}]{Reeb2014}%
  \BibitemOpen
  \bibfield  {author} {\bibinfo {author} {\bibfnamefont {D.}~\bibnamefont
  {Reeb}}\ and\ \bibinfo {author} {\bibfnamefont {M.~M.}\ \bibnamefont
  {Wolf}},\ }\href {http://stacks.iop.org/1367-2630/16/i=10/a=103011}
  {\bibfield  {journal} {\bibinfo  {journal} {New Journal of Physics}\ }\textbf
  {\bibinfo {volume} {16}},\ \bibinfo {pages} {103011} (\bibinfo {year}
  {2014})}\BibitemShut {NoStop}%
\end{thebibliography}%

\end{document}